\documentclass[11pt]{article}
\usepackage[margin=1.0in]{geometry}
\usepackage[english]{babel}
\usepackage{setspace}
\usepackage{fancyhdr}
\usepackage[T1]{fontenc}
\usepackage[utf8]{inputenc}
\usepackage{mathpazo}

\usepackage{nomencl}
\makenomenclature

\usepackage[centertags]{amsmath}
\usepackage{amsthm}
\usepackage{amssymb}
\usepackage{amsfonts}
\usepackage{bm}
\usepackage[psamsfonts,mathscr]{eucal}
\usepackage[Symbol]{upgreek}
\usepackage{mathtools}
\usepackage{siunitx}

\renewcommand{\vec}[1]{\mathbf{#1}}
\newcommand{\mat}[1]{\mathsf{#1}}

\newcommand{\bmu}{\boldsymbol{\mu}}

\newcommand{\skrom}{\texttt{scikit-rom}}

\usepackage{graphicx}
\graphicspath{{Figures/}}
\usepackage[table,xcdraw,dvipsnames]{xcolor}
\usepackage{tikz}
\usetikzlibrary{shapes.geometric,arrows.meta,positioning,fit,backgrounds,calc}
\usepackage{pgfplots}
\pgfplotsset{compat=1.18}
\usepgfplotslibrary{statistics}
\usepackage{wrapfig}
\usepackage[export]{adjustbox}
\usepackage{epsfig}
\usepackage{svg}

\definecolor{PlainBlue}{rgb}{0,0,.7}
\definecolor{PlainRed}{rgb}{.7,0,0}
\definecolor{BrownRed}{rgb}{.77,0.33,0.31}
\definecolor{ForestGreen}{HTML}{1b6f1b}
\definecolor{TUMBlue}{RGB}{0,101,189}
\definecolor{TUMOrange}{RGB}{227,114,34}
\definecolor{TUMGreen}{RGB}{162,173,0}
\definecolor{TUMRed}{RGB}{196,7,27}

\usepackage{multirow}
\usepackage{multicol}
\usepackage{booktabs}
\usepackage[most]{tcolorbox}

\usepackage{enumitem}
\usepackage{float}
\usepackage{stfloats}
\usepackage{afterpage}
\usepackage[normalem]{ulem}
\usepackage{caption}
\usepackage[labelformat=parens]{subcaption}
\usepackage[
    colorlinks=true,
    citecolor=ForestGreen,
    linkcolor=BrownRed,
    urlcolor=PlainBlue,
    bookmarks=false
]{hyperref}


\usepackage[capitalise]{cleveref}
\usepackage[numbers,sort&compress]{natbib}

\usepackage{authblk}

\usepackage{algpseudocode}

\usepackage{listings}
\renewcommand{\footnoterule}{%
    \kern -3pt
    \hrule width \textwidth height 0.5pt
    \kern 2pt%
}
\title{\texttt{scikit-rom}: An Open-Source Python Platform for Teaching and Prototyping Projection-Based Reduced-Order Modeling}
\date{}
\author[1,*]{Suparno Bhattacharyya}
\author[2]{Ali Syed}
\author[3]{Jian Tao}
\author[4]{Jean C.\ Ragusa}

\affil[1]{Department of Mechanical Engineering, Indian Institute
of Technology (ISM) Dhanbad, Jharkhand 826004, India}

\affil[2]{Department of Mathematics, Texas A\&M University,
College Station, TX 77843, USA}

\affil[3]{Department of Visualization / Institute of Data Science,
Texas A\&M University, College Station, TX 77843, USA}

\affil[4]{Department of Nuclear Engineering, Texas A\&M University,
College Station, TX 77843, USA}

\affil[*]{Corresponding author: \texttt{suparno@iitism.ac.in}}

\begin{document}
\maketitle

\begin{abstract}

Projection-based reduced-order modeling (ROM) has become a cornerstone technique for accelerating parameter-intensive engineering simulations, yet the methodology remains challenging to teach and prototype in practice. The difficulty stems from the inherently multi-stage nature of the workflow — encompassing snapshot generation, SVD/POD basis construction, Galerkin projection, offline–online decomposition, hyper-reduction, and error assessment — each of which demands careful exposition and hands-on experimentation. Existing software frameworks either abstract these stages behind opaque high-level interfaces or depend on compiled, monolithic solver stacks that are impractical for students and early-stage researchers to inspect or modify. This paper introduces \texttt{scikit-rom}, an open-source Python library designed to make the complete projection-based ROM pipeline transparent, modifiable, and amenable to interactive notebook-based exploration.
Built on the lightweight finite element backend \texttt{scikit-fem}, \texttt{scikit-rom} furnishes a coherent set of modular problem templates together with dedicated facilities for snapshot generation, reduced-basis construction, reduced-operator assembly, online ROM solution, hyper-reduction, and quantitative accuracy assessment — all within a single, unified workflow. Hyper-reduction is supported through four established strategies: the Discrete Empirical Interpolation Method (DEIM), S-OPT sampling, the Energy-Conserving Sampling and Weighting (ECSW) scheme, and ECM-style cubature construction. The capability of the framework is demonstrated through four progressive worked examples that guide the reader from full-order simulation to ROM and hyper-reduced model construction across linear, nonlinear, static, and transient problem classes. The library is primarily intended for graduate-level instruction, intensive workshops, and research prototyping in the reduced-order modeling community.
The project is maintained openly, with source code and examples available at \href{https://github.com/suparnob100/scikit-rom}{github.com/suparnob100/scikit-rom} and documentation at \href{https://scikitrom.github.io}{scikitrom.github.io}.
\end{abstract}

\section{Introduction}
\label{sec:intro}

Physics-based simulation is central to modern engineering, with the Finite Element Method (FEM) as a key numerical framework across many fields~\cite{hughes2012finite}. Increasingly, simulation is used not just for forward analysis but within iterative, data-driven pipelines that require repeated evaluations under varying parameters. This \textit{many-query} setting arises in design optimization, uncertainty quantification, Bayesian inference, and real-time digital twins~\cite{benner2015survey,ghattas2021learning}. For high-fidelity models with $N > \mathcal{O}(10^{6})$ degrees of freedom, even a single solve can be costly, making repeated queries computationally prohibitive.
 
A common solution to this problem is reduced order modeling (ROM), in which expensive high-fidelity computations are performed \textit{offline} to train a ROM, which then enables rapid \textit{online} queries. This approach, rooted in Proper Orthogonal Decomposition (POD)~\cite{Sirovich1987,Berkooz1993} and formalized by the Reduced Basis (RB) method~\cite{Quarteroni2016,Hesthaven2016,Rozza2008}, constructs a low-dimensional trial space and preassembles reduced operators. Online, new parameter instances are solved in this reduced space, often yielding orders-of-magnitude speedup and making many-query studies feasible.
 
For linear and parameter-affine problems, this strategy is highly effective. However, for nonlinear or non-affine problems, basis compression alone is insufficient \cite{bhattacharyya2020energy,Bhattacharyya2025_ijnlm}: reduced operators and residuals still require full-order assembly at each online step, which limits speedup. This bottleneck motivates hyper-reduction~\cite{farhat2020computational,Bhattacharyya2025}, which further approximates the problem so that online evaluation no longer depends on the full mesh.
 
Hyper-reduction methods for projection-based ROMs fall into two main categories \cite{Bhattacharyya2025}: approximate-then-project (e.g., DEIM~\cite{Chaturantabut2010}, EIM~\cite{Barrault2004}), which approximate nonlinear terms in the full space before projection, and project-then-approximate (e.g., GNAT~\cite{carlberg2013gnat}, ECSW~\cite{Farhat2014}, ECM~\cite{Hernandez2017}, EQP~\cite{Patera_2017_EQP}), which approximate projected reduced quantities directly via sparse sampling and weighting. Each approach offers different advantages depending on the problem structure and application.

This practical importance of projection-based ROM has led to the development of several software libraries \cite{puzyrev2019pyrom,negri_redbkit}. However, there remains a notable gap for users seeking a lightweight, Python-native environment in which the entire intrusive ROM workflow, for both linear and nonlinear problems, is straightforward to install, inspect, and modify.
For example, libROM~\cite{Choi2020} is a robust library designed for large-scale reduced-order modeling, but it is fundamentally a C++ package intended for high-performance, compiled workflows. Its Python interface, pylibROM, simply wraps the compiled code rather than providing a standalone Python implementation. As a result, installation still depends on a complex stack of dependencies, and the user experience remains closely tied to a parallel, MPI-based software environment.

In contrast, pyMOR~\cite{Milk2016} is a pure Python package and is comparatively easier to install, but it is designed with a different philosophy. Its architecture centers on abstract interfaces such as Operator, VectorArray, and Model, and it connects to external PDE solvers through bindings. While this design enhances interoperability, it also means that intrusive work is handled through adapter layers and backend bindings, rather than through a concise codebase in which reduced assembly and hyper-reduction logic are directly accessible.

A related and important software ecosystem has also been developed by Gianluigi Rozza's group and collaborators, including RBniCS, RBniCSx, and EZyRB~\cite{Rozza2024}. These libraries have made substantial contributions to the dissemination of reduced basis methods, POD-Galerkin reduction, and data-driven reduced modeling for parametrized problems. However, their design priorities differ from the objective of the present work. RBniCS and RBniCSx are closely tied to the FEniCS/FEniCSx ecosystem, so installation, version compatibility, and reproducibility remain linked to the underlying finite element backend. The RBniCSx project itself acknowledges that it is still in early development, with many features yet to be implemented. EZyRB, on the other hand, is well suited for data-driven and non-intrusive reduced modeling, but it does not aim to expose a complete intrusive finite-element-to-hyper-reduction workflow within a compact teaching-oriented codebase.

Therefore, although these tools provide valuable and widely used tools for reduced order modeling, they do not fully address the specific educational and prototyping need considered here: a lightweight, notebook-friendly, Python-native environment in which full-order assembly, projection, reduced assembly, hyper-reduction, and error assessment can be inspected and modified within a single coherent workflow.

This limitation is especially important in an educational setting, where the objective is not only to use ROM software but also to understand how the method is constructed. Projection-based ROM involves several linked stages, including snapshot collection, basis construction, Galerkin projection, offline--online decomposition, hyper-reduction, and error assessment. When these stages are distributed across non-transparent software layers, students and researchers may find it difficult to trace how projection and hyper-reduction algorithms are implemented or to modify individual components. Instructors therefore benefit from examples in which each stage is visible, executable, and modifiable within a single computational workflow.

What is still lacking, therefore, is a package that is easily installable across multiple platforms (Linux/Windows/Mac), runs smoothly in a notebook or lightweight cloud environment, and is structured so that users can directly view and modify each stage of the workflow: full-order assembly, snapshot generation, basis construction, sampling, weight computation, and reduced online evaluation. 
This is exactly the gap that \skrom{} aims to fill. Built on \texttt{scikit-fem}~\cite{skfem2020}, which is a pure-Python finite element assembly library designed to convert weak forms into sparse matrices and vectors, \skrom{} keeps both the solver and hyper-reduction workflow within its own environment. 
In essence, \texttt{scikit-fem} provides a lightweight and efficient assembly layer, while \skrom{} adds ROM and HyperROM capabilities on top through a modular problem template, a problem registry, and offline/online drivers that implement DEIM, ECSW, and ECM through a unified master-class interface. 
This results in a workflow that is explicit, self-contained, and easily modifiable by the user.

Commercial tools also offer ROM capabilities, but they often focus on deployable surrogate models derived purely from simulation or test data, rather than intrusive, projection-based hyper-reduction workflows. Platforms such as Siemens Simcenter~\cite{Siemens2023} and Ansys~\cite{AnsysTwinBuilder} use data-driven, statistical, and machine-learning methods for system simulation, design exploration, or digital twins. COMSOL~\cite{COMSOL2025ModelReduction,Bhattacharyya2025} is notable for also supporting explicit projection-based reductions, such as modal and POD-based methods. However, these environments are designed for use within proprietary software and do not expose the full algorithmic details needed for intrusive ROM development. Thus, while valuable for industrial applications, they are less suitable for studying, modifying, or benchmarking the complete ROM pipeline.

 The main contributions of this work are:
\begin{enumerate}[label=(\roman*)]
\item a Python-native, notebook-friendly platform for teaching and prototyping projection-based ROM;
\item a modular problem-template structure that separates full-order modeling, basis construction, projection, hyper-reduction, and error reporting;
\item unified implementations of Galerkin ROM, DEIM, S-OPT, ECSW, and ECM-style hyper-reduction;
\item executable learning modules that demonstrate ROM workflows for linear, nonlinear, static, and dynamic engineering problems.
\end{enumerate}
 
The remainder of the paper is organized as follows.
Section~\ref{sec:background} reviews the ROM and hyper-reduction theory implemented in \skrom{}.
Section~\ref{sec:architecture} presents the software architecture and core functionalities. 
Section~\ref{sec:workflow} describes the complete ROM-workflow. 
Section~\ref{sec:examples} demonstrates the framework on benchmark
problems. 
Section~\ref{sec:conclusions} summarizes the contributions and outlines future work.

\section{Background}
\label{sec:background}

\subsection{The Parametric Full Order Model}
\label{sec:fom}

Let $\Omega \subset \mathbb{R}^d$ be a bounded domain, $\mathcal{P} \subset \mathbb{R}^p$ a compact parameter set, and $\mathcal{I} = (0,T]$ a time interval. After spatial discretization, many parametric, time-dependent PDEs can be written as
\begin{equation}
  \mathcal{R}\bigl(
    t, \vec{u}(t;\boldsymbol{\mu}), \dot{\vec{u}}(t;\boldsymbol{\mu}),
    \ddot{\vec{u}}(t;\boldsymbol{\mu}); \boldsymbol{\mu}
  \bigr) = \vec{0},
  \qquad t \in \mathcal{I},
  \label{eq:fom_generic}
\end{equation}
where $\vec{u}(t;\boldsymbol{\mu}) \in \mathbb{R}^N$ is the discrete state vector, $N$ is the number of full-order degrees of freedom, and $\mathcal{R}$ is a parameter-dependent residual. Depending on the problem, the semi-discrete full-order model (FOM) may be first- or second-order in time:
\begin{align}
  \mat{M}(\boldsymbol{\mu})\,\dot{\vec{u}}(t;\boldsymbol{\mu})
  + \vec{g}\bigl(\vec{u}(t;\boldsymbol{\mu}),t;\boldsymbol{\mu}\bigr)
  &= \vec{F}(t;\boldsymbol{\mu}), 
  \label{eq:fom_first_order} \\
  \mat{M}(\boldsymbol{\mu})\,\ddot{\vec{u}}(t;\boldsymbol{\mu})
  + \mat{C}(\boldsymbol{\mu})\,\dot{\vec{u}}(t;\boldsymbol{\mu})
  + \vec{g}\bigl(\vec{u}(t;\boldsymbol{\mu}),t;\boldsymbol{\mu}\bigr)
  &= \vec{F}(t;\boldsymbol{\mu}).
  \label{eq:fom_second_order}
\end{align}
Here, $\mat{M}$ denotes the mass or capacity matrix, $\mat{C}$ denotes a rate-dependent damping matrix, $\vec{g}$ represents the internal residual vector, and $\vec{F}$ is the external forcing or source vector. The internal residual may include stiffness, diffusion, transport, reaction, or nonlinear effects. For a linear problem, $\vec{g}(\vec{u},t;\boldsymbol{\mu})=\mat{K}(\boldsymbol{\mu})\vec{u}$, where $\mat{K}(\boldsymbol{\mu})$ is the parameter-dependent stiffness, diffusion, or conductivity matrix. The large dimension $N$, determined by the mesh resolution, makes repeated queries expensive: assembly typically scales with the number of elements $N_e$, while linear or nonlinear solves scale with the algebraic system size.

\subsection{Snapshot Collection and Dimension Reduction via SVD}
\label{sec:svd}

Given a training parameter set $\{\boldsymbol{\mu}^{(i)}\}_{i=1}^{N_s}$, the FOM is solved for each parameter value. The state vectors at selected time instants are then centered and assembled into the snapshot matrix
\begin{equation}
  \mat{S} =
  \bigl[\,
  \widetilde{\vec{u}}^{(1,1)}
  \;\cdots\;
  \widetilde{\vec{u}}^{(i,k)}
  \;\cdots\;
  \widetilde{\vec{u}}^{(N_s,N_t)}
  \,\bigr]
  \in \mathbb{R}^{N \times N_s N_t},
  \label{eq:snapshot}
\end{equation}
where
\begin{equation}
  \widetilde{\vec{u}}^{(i,k)}
  =
  \vec{u}(t_k;\boldsymbol{\mu}^{(i)})
  -
  \vec{u}_{\mathrm{ref}} .
  \label{eq:centered_snapshot}
\end{equation}
The vector $\vec{u}_{\mathrm{ref}}\in\mathbb{R}^{N}$ is the reference state used for centering. It may be chosen as the snapshot mean, the initial state, or another prescribed baseline state.

The singular value decomposition of the snapshot matrix is written as
\begin{equation}
  \mat{S} = \widetilde{\mat{U}}\mat{\Sigma}\mat{W}^{\top},
  \label{eq:svd}
\end{equation}
where $\widetilde{\mat{U}}$ contains the left singular vectors, $\mat{\Sigma}$ contains the singular values, and $\mat{W}$ contains the right singular vectors. The columns of $\widetilde{\mat{U}}$ represent spatial modes, while the columns of $\mat{W}$ describe the corresponding temporal or parametric coefficients. The decay of $\sigma_k^2$ reflects the effective dimension of the solution manifold~\cite{Bhattacharyya2025}.

The reduced basis is obtained by retaining the first $r$ columns of $\widetilde{\mat{U}}$:
\begin{equation}
  \mat{U}
  =
  \widetilde{\mat{U}}_{(:,1:r)}
  \in \mathbb{R}^{N\times r}.
  \label{eq:reduced_basis}
\end{equation}
The reduced dimension $r$ is chosen using the cumulative variance criterion
\begin{equation}
  r = \min\left\{
  k \;\middle|\;
  \frac{\sum_{j=1}^{k}\sigma_j^2}
       {\sum_{j=1}^{N_s N_t}\sigma_j^2}
  \geq \eta
  \right\},
  \qquad \eta = 0.99.
  \label{eq:energy}
\end{equation}

\begin{figure}[t]
\centering
\begin{tikzpicture}
  \begin{axis}[
      name        = ax1,
      width       = 0.40\textwidth,
      height      = 5.0cm,
      xlabel      = {Mode index $k$},
      ylabel      = {$\sigma_k / \sigma_1$},
      ymode       = log,
      ymin        = 1e-12, ymax = 2,
      xmin        = 1,     xmax = 15,
      xtick       = {1,5,10,15},
      ytick       = {1,1e-3,1e-6,1e-9,1e-12},
      grid        = both,
      grid style  = {line width=0.3pt, draw=gray!30},
      tick label style = {font=\footnotesize},
      label style      = {font=\small},
      title       = {\small (a) Singular value decay},
      title style = {yshift=-2pt},
      legend style = {
          font=\scriptsize,
          at={(0.97,0.97)},
          anchor=north east,
          draw=none,
          fill=white,
          fill opacity=0.8
      },
  ]
    \addplot[
        color      = TUMBlue,
        line width = 1.4pt,
        mark       = *,
        mark size  = 2pt,
    ] coordinates {
        (1,  1.000e+00)
        (2,  8.500e-02)
        (3,  6.200e-03)
        (4,  4.100e-04)
        (5,  2.800e-05)
        (6,  1.500e-06)
        (7,  7.000e-08)
        (8,  3.000e-09)
        (9,  1.200e-10)
        (10, 4.000e-12)
        (11, 1.500e-12)
        (12, 1.000e-12)
        (13, 1.000e-12)
        (14, 1.000e-12)
        (15, 1.000e-12)
    };
    \addlegendentry{$\sigma_k/\sigma_1$}
  \end{axis}

  \begin{axis}[
      name        = ax2,
      at          = {($(ax1.east)+(2.8cm,0)$)},
      anchor      = west,
      width       = 0.40\textwidth,
      height      = 5.0cm,
      xlabel      = {Mode index $k$},
      ylabel      = {$\mathcal{E}(k)$},
      ymin        = 0, ymax = 1.05,
      xmin        = 1, xmax = 15,
      xtick       = {1,5,10,15},
      ytick       = {0,0.2,0.4,0.6,0.8,1.0},
      yticklabels = {0,0.2,0.4,0.6,0.8,1.0},
      grid        = both,
      grid style  = {line width=0.3pt, draw=gray!30},
      tick label style = {font=\footnotesize},
      label style      = {font=\small},
      title       = {\small (b) Cumulative variance $\mathcal{E}(k)$},
      title style = {yshift=-2pt},
      legend style = {
          font=\scriptsize,
          at={(0.05,0.60)},
          anchor=west,
          draw=none,
          fill=white,
          fill opacity=0.8
      },
  ]
    \addplot[
        color      = TUMOrange,
        line width = 1.4pt,
        mark       = *,
        mark size  = 2pt,
    ] coordinates {
        (1,  0.7841)
        (2,  0.8559)
        (3,  0.9112)
        (4,  0.9513)
        (5,  0.9741)
        (6,  0.9876)
        (7,  0.9934)
        (8,  0.9967)
        (9,  0.9983)
        (10, 0.9991)
        (11, 0.9995)
        (12, 0.9998)
        (13, 0.9999)
        (14, 1.0000)
        (15, 1.0000)
    };
    \addlegendentry{$\mathcal{E}(k)$}

    \addplot[
        color      = TUMGreen,
        dashed,
        line width = 1.2pt,
        domain     = 1:20,
    ] {0.99};
    \addlegendentry{$\eta = 0.99$}

    \addplot[
        color      = TUMRed,
        dotted,
        line width = 1.2pt,
    ] coordinates {(7, 0) (7, 1.05)};
    \addlegendentry{$r = 7$}

    \node[font=\scriptsize, TUMRed, anchor=south west]
      at (axis cs:7.15,0.05) {$r=7$};
  \end{axis}
\end{tikzpicture}
\caption{Singular value analysis of the snapshot matrix $\mat{S}$ for a representative parametric problem. \emph{Left}: normalized singular values $\sigma_k/\sigma_1$ as a function of the mode index. \emph{Right}: cumulative variance $\mathcal{E}(k)=\sum_{j=1}^{k}\sigma_j^2/\sum_{j=1}^{N_sN_t}\sigma_j^2$. The threshold $\eta=0.99$ is reached at $r=7$, which determines the reduced-basis dimension according to \eqref{eq:energy}.}
\label{fig:svd_decay}
\end{figure}
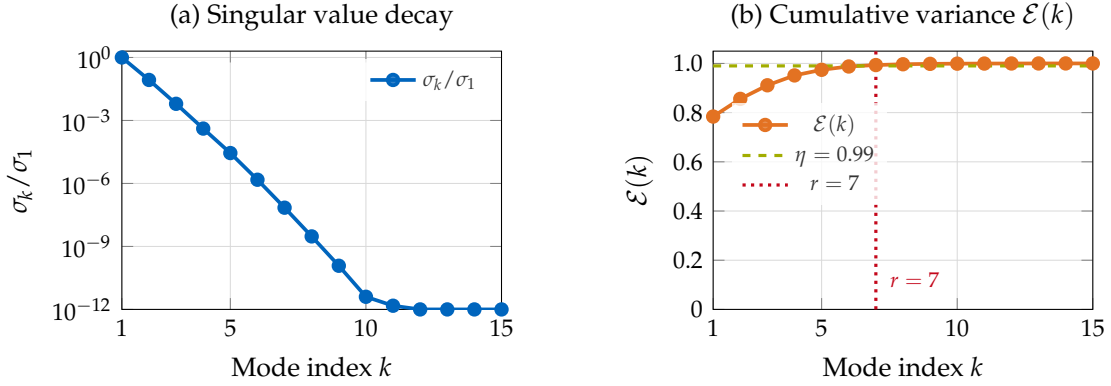

\subsection{Galerkin Projection and the Offline--Online Decomposition}
\label{sec:galerkin}

With the reduced basis $\mat{U}\in\mathbb{R}^{N\times r}$, the ROM approximation is written as
\begin{equation}
  \vec{u}(t;\bmu)
  \approx
  \vec{u}_{\textrm{ref}}
  +
  \mat{U}\widehat{\vec{u}}(t;\bmu),
  \label{eq:rom_ansatz}
\end{equation}
where $\widehat{\vec{u}}(t;\bmu)\in\mathbb{R}^{r}$ is the reduced coordinate vector. Galerkin projection yields a reduced system of dimension $r\ll N$.

For linear parameter-affine problems, the full-order matrix can be expressed as
\begin{equation}
  \mat{K}(\bmu)
  =
  \sum_{q=1}^{Q_K}
  \theta_q^K(\bmu)\,\mat{K}_q,
  \label{eq:affine_full_operator}
\end{equation}
where $\theta_q^K(\bmu)$ are parameter-dependent scalar coefficients and $\mat{K}_q$ are parameter-independent matrices. The reduced matrix is then preassembled offline as
\begin{equation}
  \widehat{\mat{K}}(\bmu)
  =
  \sum_{q=1}^{Q_K}
  \theta_q^K(\bmu)\,\widehat{\mat{K}}_q,
  \qquad
  \widehat{\mat{K}}_q
  =
  \mat{U}^{\top}\mat{K}_q\mat{U}.
  \label{eq:reduced_affine_operator}
\end{equation}
The same projection is applied to other matrices and forcing vectors. In the online stage, only reduced matrices and vectors are assembled, so the online cost is independent of the full dimension $N$ for affine linear problems.

\subsection{Hyper-reduction of Nonlinear Operators}
\label{sec:hyperreduction}

For nonlinear systems, Galerkin projection alone does not remove all full-order costs. The projected nonlinear vector
\begin{equation}
  \widehat{\vec{g}}(\widehat{\vec{u}};t,\bmu)
  =
  \mat{U}^{\top}
  \vec{g}\bigl(
    \vec{u}_{\textrm{ref}}
    +
    \mat{U}\widehat{\vec{u}},
    t;\bmu
  \bigr)
  \label{eq:projected_nonlinear_term}
\end{equation}
still requires evaluating $\vec{g}$ in the full-order space $\mathbb{R}^{N}$. Thus, even though the Newton or time-stepping system has reduced dimension $r$, the online evaluation of nonlinear residuals and Jacobians may still scale with the full mesh.

Hyper-reduction methods reduce this remaining cost by approximating nonlinear contributions using a small number of sampled entries, elements, or integration points. In \skrom{}, methods such as DEIM, ECSW, and ECM follow this principle within the same projection-based workflow. The goal is to replace the full nonlinear evaluation in \eqref{eq:projected_nonlinear_term} by an approximation whose cost depends on the number of selected samples rather than on $N$ or $N_e$.

\subsection{Error Assessment}
\label{sec:error}

ROM accuracy is assessed by comparing the reconstructed reduced solution with the corresponding FOM solution over a test set. At time $t_k$ and parameter $\bmu$, the reconstructed state is
\begin{equation}
  \vec{u}_r(t_k;\bmu)
  =
  \vec{u}_{\textrm{ref}}
  +
  \mat{U}\widehat{\vec{u}}(t_k;\bmu).
  \label{eq:rom_reconstruction}
\end{equation}
A space-time root mean square error is defined as
\begin{equation}
  \mathrm{RMSE}(\bmu)
  =
  \sqrt{
    \frac{1}{N\,N_t}
    \sum_{k=1}^{N_t}
    \left\|
      \vec{u}(t_k;\bmu)
      -
      \vec{u}_r(t_k;\bmu)
    \right\|_2^2
  }.
  \label{eq:error}
\end{equation}
The relative space-time $L^2$ error is computed as
\begin{equation}
  \epsilon_{L^2}^{\mathrm{rel}}(\bmu)
  =
  \left(
  \frac{
    \sum_{k=1}^{N_t}
    \left\|
      \vec{u}(t_k;\bmu)
      -
      \vec{u}_r(t_k;\bmu)
    \right\|_2^2
  }{
    \sum_{k=1}^{N_t}
    \left\|
      \vec{u}(t_k;\bmu)
    \right\|_2^2
  }
  \right)^{1/2}.
  \label{eq:relative_l2_error}
\end{equation}
Relative $L^\infty$ error and $R^2$ score are also used to quantify ROM accuracy and generalization over the test parameter set.

\section{Software Architecture of \texttt{scikit-rom}}
\label{sec:architecture}

The architecture of \texttt{scikit-rom} is designed to support the standard offline--online decomposition that forms the foundation of projection-based reduced-order modeling. In this regard, \skrom{} follows a similar structure to libraries such as \texttt{libROM}~\cite{Choi2020}. The core workflow of \texttt{scikit-rom} is managed primarily by two central classes: \texttt{masterclass.py} and \\ \texttt{masterclass\_parallel.py}. The former provides a unified interface for problem definition, maintains a problem registry, coordinates the offline snapshot-generation process, and manages the online evaluation of both standard ROM and HyperROM models. The latter extends this architecture to support parallel execution via multithreading for tasks like snapshot generation and ROM evaluation.

For each problem instance, a corresponding Jupyter notebook, \texttt{problem\_<name>.ipynb}, is created within its specific folder. Although this notebook is not part of the core API, it serves as the main execution environment for users, allowing them to launch the workflow, control the offline and online phases, choose the reduced basis, invoke the ROM or HyperROM solvers, and produce plots or diagnostic output. In this setup, \texttt{problem\_def.py} defines the executable problem class, while the notebook acts as a reproducible, script-like interface for running and analyzing the workflow.

Within this framework, computationally demanding tasks such as high-fidelity simulation and snapshot collection are carried out during the offline phase. During the online phase, stored data are used for reduced-order reconstruction, solution projection, error evaluation, and, where relevant, hyper-reduced computations. This workflow creates a clear separation among problem definition, full-order modeling, reduced modeling, and acceleration techniques.

\begin{figure}
    \centering
    \includegraphics[width=\textwidth, angle=0]{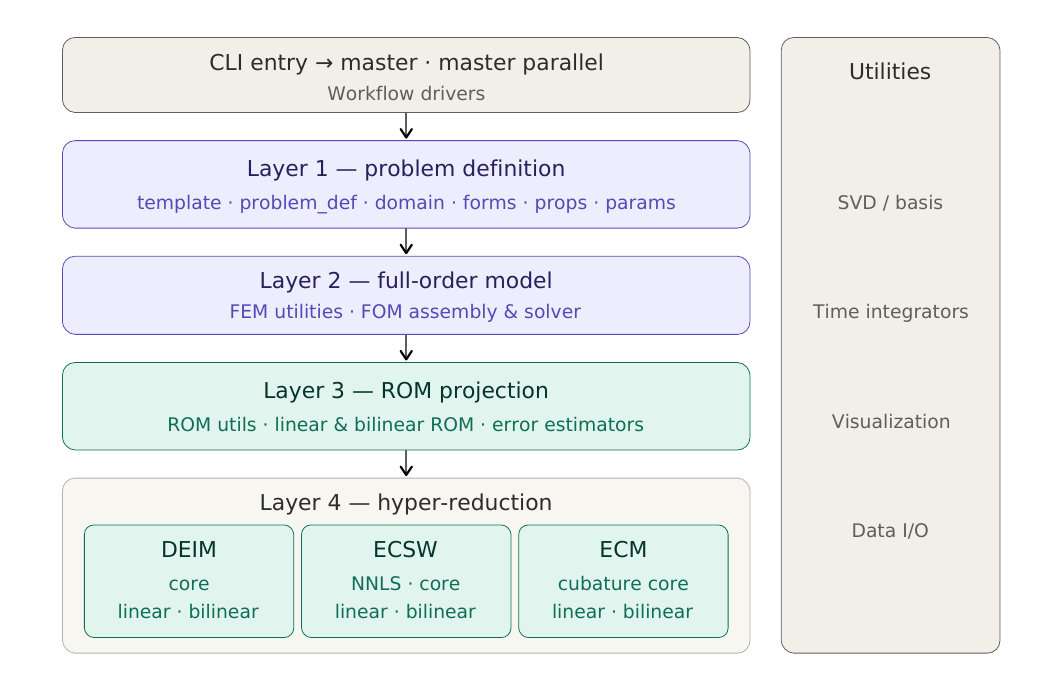}
    \caption{Software architecture of \skrom{}. The workflow is organized into problem definition, full-order simulation, ROM projection, hyper-reduction, and shared utilities for basis construction, time integration, visualization, and data storage.}
    \label{fig:pipeline}
\end{figure}

In line with this structure, \texttt{scikit-rom} is organized into four main computational layers, supported by a shared utilities module. These layers map to the principal stages of the workflow: (i) problem specification, (ii) assembly and solution of the full-order model, (iii) reduced-order projection and evaluation, and (iv) hyper-reduction for efficient online analysis. The utilities subsystem offers common services such as basis extraction, time integration, visualization, and data input/output. A schematic overview of these architectural components and their interactions is shown in~\Cref{fig:pipeline}.

\subsection{Layer 1 — Problem definition}
\label{sec:layer1}



Layer~1 defines how a user problem is introduced into \texttt{scikit-rom}. A new problem folder can be generated from the command line using
\begin{lstlisting}[language=bash,caption={Creating a new problem template.}]
skrom create-problem <name>
\end{lstlisting}
For example,
\begin{lstlisting}[language=bash]
skrom create-problem heat_2d
\end{lstlisting}
creates a folder named \texttt{problem\_heat\_2d}. The command copies the bundled problem template, renames the notebook stub to \texttt{problem\_<name>.ipynb}, and prepares the associated \texttt{problem\_def.py} file for editing.

In the recommended template, the problem is organized through six files: \texttt{domain.py}, \texttt{bilinear\_forms.py}, \texttt{linear\_forms.py}, \texttt{properties.py}, \texttt{params.py}, and \texttt{problem\_def.py}. Among these, \texttt{problem\_def.py} serves as the binding file. It registers the problem class and exposes the methods used by the serial or parallel master-class workflow. The separation into six files improves readability and reuse, but it is not mandatory. Users may define some or all components directly inside \texttt{problem\_def.py}, provided that the resulting class supplies the methods required by the workflow.

A simplified version of the generated \texttt{problem\_def.py} template is shown below. The complete template contains additional comments and docstrings, but follows the same structure.

\begin{lstlisting}[
language=Python,
caption={Compact structure of the generated \texttt{problem\_def.py} template.},
label={lst:problem_def_template},
basicstyle=\ttfamily\small
]
import os
from skrom.problem_classes.masterclass import register_problem, Problem

PROBLEM_NAME = os.path.basename(os.path.dirname(__file__))

@register_problem(PROBLEM_NAME)
class ProblemTemplate(Problem):

    # Geometry, forms, material data, and parameters
    def domain(self): ...
    def bilinear_forms(self): ...
    def linear_forms(self): ...
    def properties(self): ...
    def parameters(self, n_samples): ...

    # Full-order model
    def fom_operators(self, cls): ...
    def fom_rhs(self, cls): ...
    def fom_solver(self, cls, param): ...

    # Reduced-order model
    def reduced_operators(self, cls, param): ...
    def rom_solver(self, cls, param): ...

    # Hyper-reduced models
    def hyper_rom_operators_ecsw(self, cls, param): ...
    def hyper_rom_operators_deim(self, cls, param): ...
    def hyper_rom_operators_ecm(self, cls, param): ...
    def hyper_rom_solver_ecsw(self, cls, param): ...
    def hyper_rom_solver_deim(self, cls, param): ...
    def hyper_rom_solver_ecm(self, cls, param): ...
\end{lstlisting}

Layer~1 therefore provides a standard starting structure for new examples while allowing users to modify the generated files for problem-specific geometry, weak forms, parameters, solvers, and hyper-reduction choices. Table~\ref{tab:layer1} summarizes the primary files associated with this recommended template.

\begin{table}[ht]
\centering
\caption{Files in the Layer~1 problem template and their
architectural roles.}
\label{tab:layer1}
\resizebox{\textwidth}{!}{%
\begin{tabular}{ll}
\hline
\textbf{File} & \textbf{Role} \\
\hline
\texttt{problem\_classes/masterclass.py} &
    Abstract base class enforcing the Problem interface \\
\texttt{templates/problem\_template/domain.py} &
    Mesh construction and basis scaffold \\
\texttt{templates/problem\_template/bilinear\_forms.py} &
    Stiffness and mass form scaffold \\
\texttt{templates/problem\_template/linear\_forms.py} &
    Load-vector form scaffold \\
\texttt{templates/problem\_template/properties.py} &
    Material-property scaffold \\
\texttt{templates/problem\_template/params.py} &
    Parameter-space and sampling scaffold \\
\texttt{templates/problem\_template/problem\_def.py} &
    Concrete class binding all components \\
\hline
\end{tabular}
}
\end{table}




\subsection{Layer 2 — Full-order model}
\label{sec:layer2}

Layer~2 assembles and solves the high-fidelity finite-element model from which the reduced model is constructed. In the standard workflow, the master-class driver loops over the sampled parameter values and calls the full-order routines exposed by \texttt{problem\_def.py}. These routines may include \texttt{fom\_operators(cls)} for assembling or preparing full-order operators, \texttt{fom\_rhs(cls)} for defining the corresponding right-hand side, and \texttt{fom\_solver} for solving the full-order problem at a given parameter value.

This structure keeps the full-order model problem-specific while preserving a common workflow for snapshot generation, timing, data storage, and train--test organization. Users can therefore implement static or transient solvers, customized forcing, time grids, boundary-condition handling, nonlinear iterations, or auxiliary assembly steps inside \texttt{problem\_def.py} without modifying the surrounding library. The solution data and operators generated at this stage provide the inputs used later for basis construction, ROM projection, and, when applicable, hyper-reduction.




\subsection{Layer 3 — ROM projection}
\label{sec:layer3}

Layer~3 constructs and evaluates the reduced-order model using the full-order data generated in Layer~2. The reduced basis $\mat{U}\in\mathbb{R}^{N\times r}$, obtained from the snapshot data as described earlier, defines the low-dimensional trial space used for projection. In the problem template, the ROM-specific behavior is exposed through \texttt{reduced\_operators} and \texttt{rom\_solver} in \texttt{problem\_def.py}. The first routine prepares the projected matrices and vectors, while the second solves the reduced system for a given parameter value.

For a linear full-order matrix $\mat{K}\in\mathbb{R}^{N\times N}$, Galerkin projection gives the reduced matrix
\begin{equation}
  \widehat{\mat{K}}
  =
  \mat{U}^{\top}\mat{K}\mat{U}
  \in \mathbb{R}^{r\times r}.
  \label{eq:Kr}
\end{equation}
Similarly, an $N$-dimensional full-order vector $\vec{f}\in\mathbb{R}^{N}$ is projected as
\begin{equation}
  \widehat{\vec{f}}
  =
  \mat{U}^{\top}\vec{f}
  \in \mathbb{R}^{r}.
\end{equation}
The online problem is therefore solved in a reduced space of dimension $r\ll N$.

After the reduced coordinate vector $\widehat{\vec{u}}\in\mathbb{R}^{r}$ is computed, the corresponding full-order approximation can be reconstructed as
\begin{equation}
  \vec{u}_{r}
  =
  \vec{u}_{\mathrm{ref}}
  +
  \mat{U}\widehat{\vec{u}},
\end{equation}
where $\vec{u}_{\mathrm{ref}}\in\mathbb{R}^{N}$ is the reference state used during snapshot centering. This reconstructed field can then be used for visualization, error estimation, or comparison with stored full-order solutions. Layer~3 also includes utilities for ROM data handling, parameter sampling, and static or time-dependent accuracy reporting. \Cref{tab:layer3} summarizes the main modules included in this layer.

\begin{table}[ht]
\centering
\caption{Modules in Layer~3 (ROM projection) and their roles.}
\label{tab:layer3}
\resizebox{\textwidth}{!}{%
\begin{tabular}{ll}
\hline
\textbf{Module} & \textbf{Role} \\
\hline
\texttt{bilinear\_form\_rom.py} &
    Galerkin projection of bilinear-form matrices \\
\texttt{linear\_form\_rom.py} &
    Projection of load vectors onto the reduced basis \\
\texttt{rom\_utils.py} &
    Snapshot handling, parameter sampling, reconstruction,
    and ROM nonlinear solvers \\
\texttt{rom\_error\_est.py} &
    Static error metrics and diagnostic comparisons \\
\texttt{rom\_error\_est\_t.py} &
    Time-dependent error metrics for transient ROMs \\
\hline
\end{tabular}
}
\end{table}

\subsection{Layer 4 — Hyper-reduction}
\label{sec:layer4}

While Galerkin projection reduces the dimension of the algebraic system, it does not always remove the dominant online cost. For nonlinear or non-affine problems, reduced residuals and Jacobians may still require assembly over the full mesh. Layer~4 addresses this bottleneck by replacing full-mesh evaluation with contributions from selected degrees of freedom, elements, or quadrature points.

In the problem template, hyper-reduction is exposed through the HyperROM routines in \texttt{problem\_def.py}. These include methods for preparing method-specific operators, such as \texttt{hyper\_rom\_operators\_ecsw}, \texttt{hyper\_rom\_operators\_ecm} and \texttt{hyper\_rom\_operators\_deim}, together with solver methods such as \texttt{hyper\_rom\_solver\_ecsw}, \texttt{hyper\_rom\_solver\_deim}, and \texttt{hyper\_rom\_solver\_ecm}. This mirrors the ROM structure of Layer~3, so users can move from standard projection to hyper-reduced evaluation without changing the surrounding workflow.

Four hyper-reduction methods are currently implemented: the Discrete Empirical Interpolation Method (DEIM), Energy Conserving Sampling and Weighting (ECSW), Empirical Cubature Method (ECM), and S-OPT. These methods differ in their sampling units and approximation strategies: DEIM and S-OPT select interpolation degrees of freedom, ECSW selects weighted elements, and ECM selects weighted integration points. Their common purpose is to reduce the online cost of evaluating nonlinear or non-affine terms while retaining acceptable accuracy. Their intended use cases are summarized in Table~\ref{tab:layer4} and discussed in detail in the next section.

\begin{table}[ht]
\centering
\caption{Hyper-reduction methods in Layer~4 and their intended use
cases.}
\label{tab:layer4}
\resizebox{\textwidth}{!}{%
\begin{tabular}{lll}
\hline
\textbf{Method} & \textbf{Directory} & \textbf{Use case} \\
\hline
DEIM & \texttt{rom/deim/} &
    Nonlinear residual or force approximation by point selection \\
ECSW & \texttt{rom/ecsw/} &
    Element-level integration with energy-aware weighting \\
ECM & \texttt{rom/ecm/} &
    Constrained cubature-based element selection \\
\hline
\end{tabular}
}
\end{table}



\subsection{Utilities subsystem}
\label{sec:utilities}

The utilities subsystem provides shared support for basis computation, time integration, visualization, and data I/O. SVD-based routines are used for reduced-basis construction, while the dynamics module provides time integrators such as Newmark and WBZ-$\alpha$ for transient FOM and ROM simulations~\cite{newmark1959method,wood1980alpha}. Visualization utilities include plotting helpers and export routines for VTK and VTU formats~\cite{schroeder2006visualization}.

For data storage, the main ROM workflow writes offline data to the \texttt{ROM\_data} directory using \texttt{.npy} and compressed \texttt{.npz} files. Additional HDF5 utilities are available for array-list storage with metadata. The main modules are listed in \cref{tab:utils}.

\begin{table}[ht]
\centering
\caption{Modules in the Utilities subsystem.}
\label{tab:utils}
\resizebox{\textwidth}{!}{%
\begin{tabular}{ll}
\hline
\textbf{Module} & \textbf{Role} \\
\hline
\texttt{utils/imports.py} &
    Shared numerical and finite-element imports \\
\texttt{utils/reduced\_basis/svd.py} &
    SVD/POD basis extraction and mode selection \\
\texttt{utils/dynamics/integrators.py} &
    Time integrators for FOM and ROM simulations \\
\texttt{utils/dynamics/newmark\_beta.py} &
    Standalone Newmark-\(\beta\) implementation \\
\texttt{utils/visualization/generate\_vtk.py} &
    VTK export and time-series output \\
\texttt{utils/visualization/generate\_vtu.py} &
    VTU file generation \\
\texttt{utils/visualization/vtuwriter.py} &
    Low-level VTU writing support \\
\texttt{utils/visualization/color\_palette.py} &
    Plot styling and colour management \\
\texttt{utils/visualization/plot\_utils.py} &
    Diagnostic and comparison plotting helpers \\
\texttt{utils/hdf5\_store.py} &
    HDF5-based storage of snapshots and ROM artifacts \\
\texttt{utils/save\_h5.py} &
    Alternative entry point to HDF5 storage \\
\texttt{utils/data\_io/} &
    Additional data I/O helpers \\
\hline
\end{tabular}
}
\end{table}

\section{Software Workflow in \texttt{scikit-rom}}
\label{sec:workflow}

Having outlined the overall architecture, we now provide an overview of the typical execution flow in a \texttt{scikit-rom} session.

\subsection{Standard ROM Workflow}

A standard workflow begins with parameter sampling and splitting the high-fidelity data into training and test sets. The library offers several sampling strategies, including random permutation, Latin hypercube sampling~\cite{mckay1979comparison}, and Sobol sequences~\cite{sobol1967distribution}. Each method produces Boolean masks that specify which samples are allocated for basis construction and which are set aside for validation.

Given a training parameter $\bmu$, let $\mat{W}_{\mu}$ represent the corresponding snapshot block, such as a time-history matrix. Prior to constructing the basis, this snapshot block is centered with respect to a reference state $\vec{u}_{\mathrm{ref}}$:
\begin{equation}
  \widetilde{\mat{W}}_{\mu}
  =
  \mat{W}_{\mu}
  -
  \vec{u}_{\mathrm{ref}}\vec{1}^{\top},
\end{equation}
where $\vec{1}$ is a vector of ones whose length equals the number of snapshots in $\mat{W}_{\mu}$. The reference state can be chosen either as the empirical mean of the training dataset or as the initial condition.

If the training data are already assembled into a single snapshot matrix, the reduced basis $\mat{U}$ can be obtained by applying SVD directly to this matrix. Alternatively, \texttt{scikit-rom} supports \textit{incremental basis construction} across the training parameters. For each centered snapshot block, the components already captured by the current basis $\mat{U}$ are projected out:
\begin{equation}
  \widetilde{\mat{W}}_{\perp}
  =
  \widetilde{\mat{W}}_{\mu}
  -
  \mat{U}
  \left(
  \mat{U}^{\top}\widetilde{\mat{W}}_{\mu}
  \right).
\end{equation}
An SVD is then performed on $\widetilde{\mat{W}}_{\perp}$ to identify additional directions that are not represented by the current basis. Let $\Delta\mat{U}$ denote the left singular-vector matrix obtained from this local SVD. The first $k$ enrichment modes, $\Delta\mat{U}_{(:,1:k)}$, are appended to the existing basis. The combined matrix is then re-orthonormalized using a QR factorization:
\begin{equation}
  \mat{Q}\mat{R}
  =
  \operatorname{qr}
  \left(
  \left[
  \mat{U},\;
  \Delta\mat{U}_{(:,1:k)}
  \right]
  \right),
\end{equation}
and the updated basis is taken as
\begin{equation}
  \mat{U}_{\mathrm{new}}
  =
  \mat{Q}.
\end{equation}
Here, $k$ is the number of enrichment modes retained from the current snapshot block. Thus, the reduced space can be built either from a global snapshot matrix or incrementally, using one parameter block at a time.

A key feature of \texttt{scikit-rom} is the direct construction of reduced operators from local finite-element data. For large-scale nonlinear problems, forming the global full-order matrix or vector first and projecting it afterwards is often prohibitively expensive, since the bulk of the computational effort is already spent before any reduction takes place. To circumvent this issue, \texttt{BilinearFormROM} and \texttt{LinearFormROM} operate directly on element-level contributions. Local matrices or vectors are extracted, restricted to the active local degrees of freedom, and projected onto the reduced basis prior to global accumulation. Consequently, the corresponding global full-order operator is never assembled solely for the purpose of projection.

To handle boundary constraints efficiently, the elements are split into two groups. Group~A comprises elements all of whose local degrees of freedom are unconstrained; their projected contributions can therefore be processed in batches. Group~B comprises elements that involve at least one Dirichlet-constrained degree of freedom and therefore require masked extraction at the local level. A local-to-reduced index map is precomputed once, eliminating repeated index look-ups during assembly.

For Group~A, the projected contributions are accumulated in chunks so as to bound peak memory consumption while still leveraging dense linear algebra routines. In the bilinear-form case, the central tensor contraction is
\begin{lstlisting}[language=Python]
sum_A += np.einsum(
    'eia,eij,ejb->ab',
    R_test_chunk,    # (chunk_size, n_loc, r)
    K_local_chunk,   # (chunk_size, n_loc, n_loc)
    R_trial_chunk,   # (chunk_size, n_loc, r)
    optimize=True,
)
\end{lstlisting}
which computes
\begin{equation}
  \sum_{e \in \mathcal{E}_{A}^{(c)}}
  \left(\mat{U}_{\mathrm{test}}^{e}\right)^{\top}
  \mat{K}_{e}
  \mat{U}_{\mathrm{trial}}^{e}
\end{equation}
over the current chunk $\mathcal{E}_{A}^{(c)}$. Here, $\mat{K}_{e}$ is the element-level matrix, while $\mat{U}_{\mathrm{test}}^{e}$ and $\mat{U}_{\mathrm{trial}}^{e}$ denote the restrictions of the test and trial reduced bases to the local degrees of freedom of element $e$. The same principle is applied by \texttt{LinearFormROM} to local vectors. This chunked design keeps memory usage under control while preserving the performance benefits of BLAS-backed tensor operations.

Group~B is handled on a per-element basis because the presence of constrained local degrees of freedom prevents uniform batching. Each local contribution is first restricted to its free subset and then projected individually. Although this element-by-element path is less efficient, elements of this type are typically concentrated near the boundary, so the overhead remains modest.

The operator layer also supports hyper-reduction without requiring a separate assembly workflow. In addition to returning fully accumulated reduced operators, the same form infrastructure can retain local projected contributions before final summation. This is useful because different hyper-reduction methods require access to different levels of local information. ECSW uses element-wise reduced contributions to construct a weighted sparse element sum, while ECM applies the same idea at the integration-point level. DEIM and S-OPT instead use restricted evaluations associated with selected interpolation degrees of freedom and recover the reduced nonlinear contribution through an interpolation operator. Thus, standard ROM and HyperROM share the same reduced-assembly infrastructure; they differ only in how the local projected contributions are sampled, reconstructed, or accumulated during online evaluation.

At the solution stage, the online workflow supports two modes. In the evaluation mode, the ROM is run over the reserved test parameters, reconstructed in the full-order space when needed, and compared with the stored FOM solutions to report error and speed-up. In the prediction mode, the same driver accepts a user-specified parameter list and returns the computed ROM solutions without requiring reference FOM data.

The \texttt{full\_order} option controls the representation of the stored output, not the reduced solve itself. When \texttt{full\_order=True}, the driver stores the reconstructed field
\begin{equation}
  \vec{u}_{\mathrm{ROM}}
  \approx
  \vec{u}_{\mathrm{ref}}
  +
  \mat{U}_{\mathrm{sel}}\widehat{\vec{u}},
\end{equation}
where $\widehat{\vec{u}}$ denotes the reduced coordinate vector and $\mat{U}_{\mathrm{sel}}$ denotes the selected reduced basis used in the online solve. When \texttt{full\_order=False}, the driver may store the reduced coordinates directly. In evaluation mode, reconstruction can still be performed internally so that error metrics can be computed against the FOM reference solution. The same interface is used for the standard ROM and for the ECSW-, ECM-, and DEIM-based HyperROM solvers; the difference lies in how the reduced operators are evaluated.

The offline data produced during the workflow can be stored via either a lightweight NumPy/NPZ backend or an HDF5 backend. The NumPy backend writes the full-order solutions to a standalone \texttt{.npy} file and packs the remaining metadata---basis, masks, reference state---into a compressed \texttt{.npz} archive. For larger or irregularly shaped arrays, \texttt{hdf5\_store.py} offers an \texttt{h5py}-based interface with optional compression and chunking. These two storage options allow the same workflow to scale from small tutorial notebooks to substantial simulation datasets.

\subsection{Hyper-reduction methods}

Beyond the standard projection-based ROM pipeline, \texttt{scikit-rom} provides several hyper-reduction techniques. All of them operate on the same reduced operators and differ in how nonlinear contributions are sampled, reconstructed, or reweighted. Hyper-reduction is thus treated as an additional approximation layer placed between the reduced operator definition and the online Newton solve.

\paragraph{\textbf{Energy-Conserving Sampling and Weighting}}

ECSW~\cite{farhat2014dimensional,farhat2015structure-preserving,Zahr2015,grimberg2021mesh,Bhattacharyya2025} is a project-then-approximate hyper-reduction method that constructs a reduced mesh for efficient evaluation of projected nonlinear terms. Its goal is to replace the full element summation in the reduced nonlinear term by a weighted sum over a small subset of elements while preserving the reduced quantity over the training set. In \skrom{}, ECSW is supported through the offline \texttt{hyperreduce} routine, the custom NNLS solver used to compute sparse nonnegative weights, and the classes \texttt{BilinearFormHYPERROM\_ecsw} and \texttt{LinearFormHYPERROM\_ecsw}, which assemble weighted reduced bilinear and linear contributions over the selected elements.

The offline stage begins by assembling the matrix
\begin{equation}
  \mat{G} \in \mathbb{R}^{(N_s N_t r)\times N_e}.
\end{equation}
For each training snapshot $(i,k)$, the centered solution is projected onto the reduced basis, reconstructed in the full space, and the element-level reduced nonlinear-force contributions
\begin{equation}
  \mat{U}_e^{\top}
  \vec{f}^{\,e}_{\mathrm{int}}(\bmu^{(i)},t_k)
\end{equation}
are evaluated. Here, $\mat{U}_e$ denotes the restriction of the reduced basis $\mat{U}$ to the local degrees of freedom associated with element $e$, and $\vec{f}^{\,e}_{\mathrm{int}}$ is the element-level internal-force vector. Collecting these contributions over all training parameters, time instances, and reduced basis directions yields the columns of $\mat{G}$.

Because $\mat{G}$ can be very large, \skrom{} first compresses it using an SVD-based reduction, with \texttt{scikit-learn}'s \texttt{randomized\_svd} used for large matrices:
\begin{equation}
  \mat{G}
  \approx
  \mat{\Phi}_q \mat{\Sigma}_q \mat{W}_q^{\top},
\end{equation}
where $\mat{W}_q \in \mathbb{R}^{N_e \times q}$ contains the retained right singular vectors based on the singular value decay. The compressed target corresponding to uniform element weighting is then
\begin{equation}
  \vec{d}_q
  =
  \mat{W}_q^{\top}\vec{1},
\end{equation}
where $\vec{1}\in\mathbb{R}^{N_e}$ is the vector of unit element weights. This step reduces the size of the ECSW training problem before sparse element selection is performed.

The sparse nonnegative element weights are then computed by the custom NNLS algorithm adapted from \texttt{libROM}~\cite{Choi2020}. This step identifies the reduced mesh itself, namely the selected elements and their weights. Thus, randomized SVD performs the compression, whereas NNLS performs the sparse selection.

After training, the online hyper-reduced nonlinear term is assembled using only the selected elements:
\begin{equation}
  \widehat{\vec{f}}_{\mathrm{int}}^{\mathrm{ECSW}}
  (\widehat{\vec{u}};\bmu)
  =
  \sum_{e\,\in\,\widetilde{\mathcal{E}}}
  \xi_e\,
  \mat{U}_e^{\top}
  \vec{f}^{\,e}_{\mathrm{int}}
  \left(
  \mat{U}_e\widehat{\vec{u}}
  +
  \vec{u}_{\textrm{ref}}^{\,e};
  \bmu
  \right),
  \label{eq:ecsw_online}
\end{equation}
where $\widetilde{\mathcal{E}}=\{e:\xi_e>0\}$ is the reduced element set, $\xi_e$ is the ECSW weight for element $e$, and $\vec{u}_{\textrm{ref}}^{\,e}$ is the restriction of the reference state to the local degrees of freedom of element $e$. In the implementation, this restricted assembly is enabled through \texttt{scikit-fem}'s \texttt{with\_element} functionality, which constructs a finite-element basis on the selected element subset. Since $|\widetilde{\mathcal{E}}| \ll N_e$ in practice, the online cost is reduced accordingly.

\paragraph{\textbf{Empirical Cubature Method}}

ECM~\cite{hernandez2024cecm,Hernandez2017,Bhattacharyya2025} extends the sparse approximation principle to the quadrature level. Rather than selecting whole elements, it constructs a sparse cubature rule over element--Gauss-point contributions. If each element contains $N_g$ Gauss points, the complete integration rule comprises $N_eN_g$ points. ECM identifies a small subset of these points together with associated positive weights such that the dominant projected nonlinear contributions are reproduced with sufficient accuracy. Compared with ECSW, ECM offers finer-grained control over the sparsity pattern because the selection operates below the element level.

In \skrom{}, ECM follows the same HyperROM form philosophy as ECSW, but with weights assigned at the element--Gauss-point level. The bilinear and linear ECM form classes, \texttt{BilinearFormHYPERROM\_ecm} and \texttt{LinearFormHYPERROM\_ecm}, are responsible for assembling quadrature-weighted reduced matrices and vectors. As with ECSW, restricted assembly relies on \texttt{scikit-fem}'s \texttt{with\_element} mechanism to evaluate only the active elements, while the ECM weights determine which Gauss-point contributions are retained inside those elements.

\paragraph{\textbf{Discrete Empirical Interpolation Method}}

DEIM~\cite{Chaturantabut2010,tiso2013discrete,Barrault2004,Bhattacharyya2025} follows an approximate-then-project strategy. Instead of approximating only the reduced solution space, DEIM constructs a separate empirical basis for the nonlinear vector and then selects a small set of interpolation degrees of freedom. In \skrom{}, the offline DEIM selection is implemented through the \texttt{deim} class. This class builds the nonlinear snapshot basis, selects interpolation rows, maps the selected degrees of freedom to active finite elements, and constructs the interpolation matrix used during online evaluation. The online reduced assembly is then performed by \texttt{BilinearFormHYPERROM\_deim} and \texttt{LinearFormHYPERROM\_deim}, which evaluate only the DEIM-selected element contributions and reconstruct the reduced nonlinear contribution through the interpolation operator.

Let $\mat{\Phi}_f\in\mathbb{R}^{N_{\mathrm{dof}}\times n_f}$ denote the truncated left singular-vector basis of the nonlinear snapshot matrix, where $n_f$ is the number of retained nonlinear modes. Let $\mathcal{I}_{\mathrm{DEIM}}$ denote the selected interpolation indices. The restriction of $\mat{\Phi}_f$ to these indices is
\begin{equation}
  \widehat{\mat{\Phi}}_f
  =
  \mat{\Phi}_f[\mathcal{I}_{\mathrm{DEIM}},:].
\end{equation}
The reduced nonlinear vector is then approximated as
\begin{equation}
  \widehat{\vec{f}}_{\mathrm{nl}}^{\,r}
  =
  \mat{D}\,
  \vec{f}_{\mathrm{nl}}[\mathcal{I}_{\mathrm{DEIM}}],
  \qquad
  \mat{D}
  =
  \mat{U}_{\mathrm{sel}}^{\top}
  \mat{\Phi}_f
  \widehat{\mat{\Phi}}_f^{+}.
  \label{eq:deim_reconstruction}
\end{equation}
Here, $\mat{U}_{\mathrm{sel}}$ is the selected ROM basis used in the online solve, $\widehat{\mat{\Phi}}_f^{+}$ is the Moore--Penrose pseudoinverse of $\widehat{\mat{\Phi}}_f$, and $\vec{f}_{\mathrm{nl}}[\mathcal{I}_{\mathrm{DEIM}}]$ contains the nonlinear vector entries evaluated only at the selected interpolation indices. Thus, the nonlinear function is not evaluated over the full set of degrees of freedom. Instead, the complete reduced nonlinear vector is recovered algebraically through $\mat{D}$. The selected degrees of freedom are also mapped back to active finite elements, so that finite-element assembly can be restricted to the element subset required by the interpolation indices.

\paragraph{\textbf{S-Optimality (S-OPT)}}

S-Optimality (S-OPT)~\cite{shin2016SOPT,grimberg2021mesh} uses the same nonlinear snapshot basis and the same DEIM-style reconstruction framework, but it changes the criterion used to select the interpolation indices. Classical DEIM selects indices through a greedy residual-based procedure. S-OPT instead chooses the interpolation indices by optimizing a matrix quality measure associated with the selected rows of the nonlinear basis. In this sense, S-OPT modifies only the offline sampling step; the online reconstruction formula remains the DEIM-type approximation in \eqref{eq:deim_reconstruction}.

Let $\mat{\Phi}_f\in\mathbb{R}^{N_{\mathrm{dof}}\times n_f}$ be the retained nonlinear basis. S-OPT first considers the thin QR factorization
\begin{equation}
  \mat{\Phi}_f
  =
  \mat{Q}_f\mat{R},
\end{equation}
where $\mat{Q}_f\in\mathbb{R}^{N_{\mathrm{dof}}\times n_f}$ has orthonormal columns. A binary selection matrix
$\mat{Z}\in\{0,1\}^{N_{\mathrm{dof}}\times n_f}$ encodes a candidate set of $n_f$ interpolation indices. The selected submatrix is
\begin{equation}
  \mat{P}
  =
  \mat{Z}^{\top}\mat{Q}_f
  \in\mathbb{R}^{n_f\times n_f}.
\end{equation}
The admissible set of selection matrices is
\begin{equation}
  \mathcal{Z}_{n_f}
  =
  \left\{
  \mat{Z}\in\{0,1\}^{N_{\mathrm{dof}}\times n_f}
  \;:\;
  \mat{Z}^{\top}\vec{1}=\vec{1},
  \quad
  \mat{Z}\vec{1}\leq \vec{1}
  \right\}.
\end{equation}
The first condition ensures that each column of $\mat{Z}$ selects one degree of freedom, while the second prevents repeated selection of the same degree of freedom.

The S-OPT interpolation indices are obtained by solving
\begin{equation}
  \mat{Z}^{*}_{\mathrm{S\mbox{-}OPT}}
  =
  \arg\max_{\mat{Z}\in\mathcal{Z}_{n_f}}
  \mathscr{S}\left(\mat{Z}^{\top}\mat{Q}_f\right),
  \label{eq:sopt_problem}
\end{equation}
where
\begin{equation}
  \mathscr{S}(\mat{P})
  =
  \left(
      \frac{
      \sqrt{
      \left|
      \det\left(\mat{P}^{\top}\mat{P}\right)
      \right|}
      }
      {
      \prod_{i=1}^{n_f}
      \left\|\mat{P}_{(:,i)}\right\|_2
      }
  \right)^{1/n_f}.
  \label{eq:sopt_score}
\end{equation}
The score $\mathscr{S}$ favors selected submatrices whose columns are well conditioned and less collinear. After the optimal selection matrix $\mat{Z}^{*}_{\mathrm{S\mbox{-}OPT}}$ is obtained, the corresponding interpolation indices define $\mathcal{I}_{\mathrm{DEIM}}$ in \eqref{eq:deim_reconstruction}. Therefore, S-OPT changes how the sampled degrees of freedom are chosen, but the online reduced nonlinear vector is still recovered using the same DEIM-type interpolation operator.

In \texttt{scikit-rom}, S-OPT is invoked through the same \texttt{deim} class by setting the \texttt{sopt} flag:
\begin{lstlisting}[language=Python, caption={S-OPT offline training.}]
sopt_class = deim_module.deim(
    prob.mesh, F_nl, V_sel, tol_f=1e-7, extra_modes=0
)
deim_mat, sampled_rows = sopt_class.select_elems(sopt=True)
\end{lstlisting}
Here, the code variable \texttt{V\_sel} corresponds to the selected ROM basis $\mat{U}_{\mathrm{sel}}$ in the mathematical notation.

\subsection{ROM Accuracy Reporting}

\texttt{scikit-rom} evaluates the predictive quality of reduced-order models (ROMs) using a diverse set of global and sample-dependent error metrics. All accuracy measures are assessed after reconstructing the ROM solutions back into the original high-dimensional (full-order) space.


Global error metrics such as relative $L^2$ and $L^\infty$ errors, RMSE, MAE, $R^2$ score, and explained variance are computed by flattening the reconstructed and reference datasets over all snapshots, time steps, and spatial degrees of freedom.

For datasets comprising multiple samples, \texttt{scikit-rom} computes local relative errors by applying the Euclidean norm across the spatial degrees of freedom for each individual sampled state. When handling time-dependent data arranged as sample--time arrays, these local errors are further resolved at each time step within every sample and a box plot of the temporal error distribution is plotted for every parameter.

A sample summary of implemented accuracy metrics is provided in \cref{lst:rom_accuracy_metrics}. Examples of these diagnostics can be found in the numerical results section.


\begin{lstlisting}[language={},caption={Example ROM run and accuracy report.},label={lst:rom_accuracy_metrics}]
===================
ROM Accuracy Report
===================
   <Global Errors> 
L2 Error:                        2.3086e+00
Relative L2 Error:               6.1930e-03
Linf Error:                      8.6686e-03
Relative Linf Error:             6.3260e-03
RMSE:                            3.2913e-04
MAE:                             7.7472e-05
    <Statistical Fit>
R2 Score:                        1.0000
Explained Variance:              1.0000
    <Error Distribution (absolute)> 
Median Abs Error:                5.8951e-06
95th Percentile Abs Error:       3.2222e-04
   <Time/Parameter-Dependent Errors> 
Average Relative L2 over samples:        5.1271e-03
Max Relative L2 over samples:            9.4107e-02
Min Relative L2 over samples:            0.0000e+00
------------------------------------------------------------
\end{lstlisting}

\subsection{Interactive tools for educational ROM exploration}
\label{sec:interactive_rom_tools}

To enhance its educational capabilities, \texttt{skrom} incorporates interactive notebook tools designed to facilitate exploration of key decisions within a projection-based reduced-order modeling (ROM) workflow. These tools enable users to interactively analyze ROM behavior and are well-suited for classroom demonstrations, tutorials, and self-directed learning.

\begin{figure}[t]
    \centering
    \includegraphics[width=0.8\linewidth]{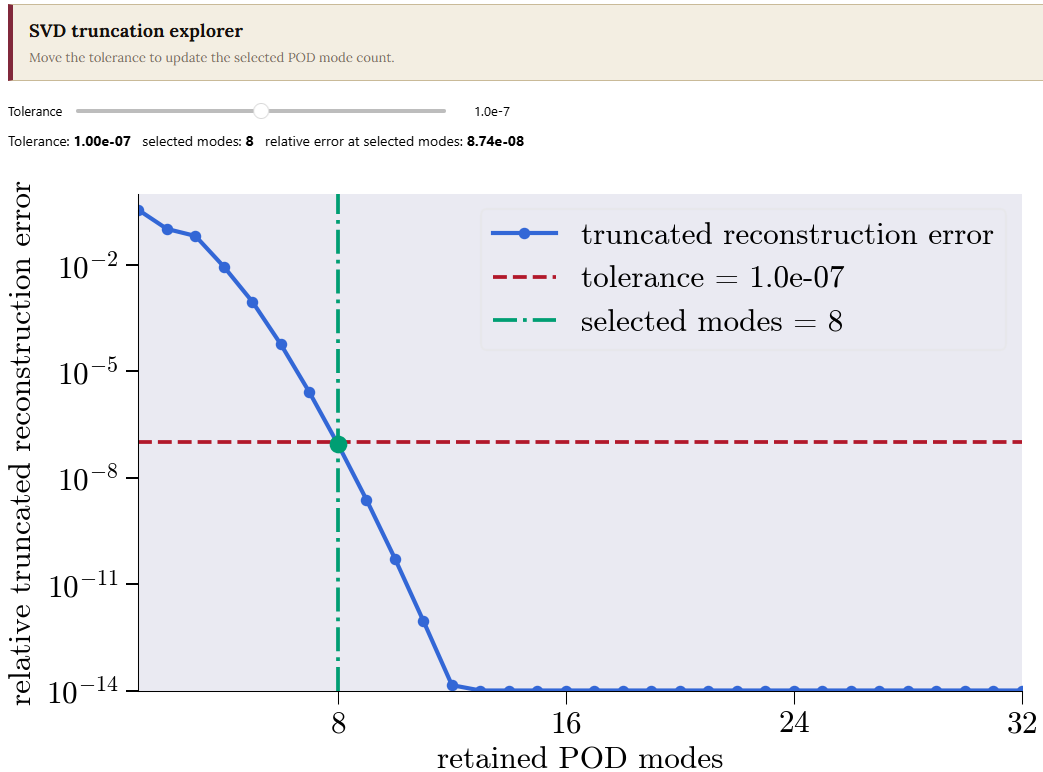}
    \caption{Interactive SVD truncation explorer in \skrom{}, showing how the tolerance determines the selected POD basis size.}
    \label{fig:interactive_svd}
\end{figure}

The first tool, the \texttt{SVD truncation explorer}, as shown in \cref{fig:interactive_svd}, visualizes the relative truncated reconstruction error as a function of the number of retained Proper Orthogonal Decomposition (POD) modes. By modifying the truncation tolerance, users can observe how the reduced dimension changes accordingly. This creates a direct visual connection between the singular value decay, the truncation threshold, and the dimension of the reduced basis.

\begin{figure}[t]
    \centering
    \includegraphics[width=\linewidth]{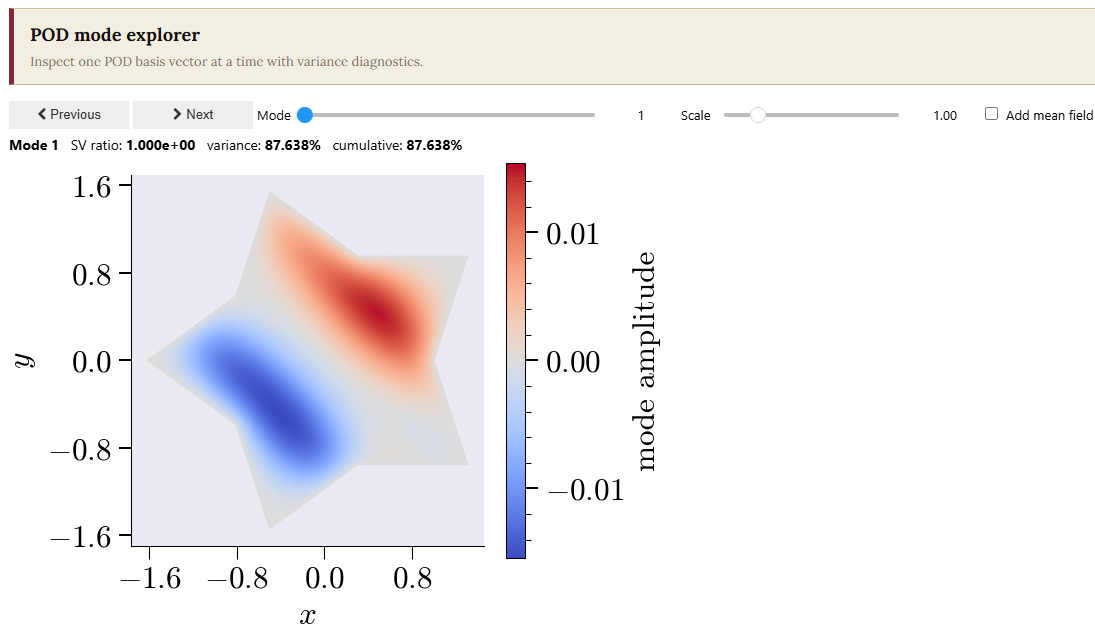}
    \caption{Interactive POD mode explorer in \skrom{} for inspecting individual basis vectors and their variance contribution.}
    \label{fig:interactive_pod}
\end{figure}

The second tool, the \texttt{POD mode explorer}, shown in \cref{fig:interactive_pod} enables users to examine each POD basis vector individually, displaying not only the vector itself, but also its singular value ratio, variance contribution, and cumulative variance. Users may rescale the mode amplitude and, when available, overlay the reference field. This feature helps interpret the POD basis as a set of spatial structures extracted from the full-order data, rather than as abstract vectors from the SVD algorithm.

\begin{figure}[t]
    \centering
    \includegraphics[width=\linewidth]{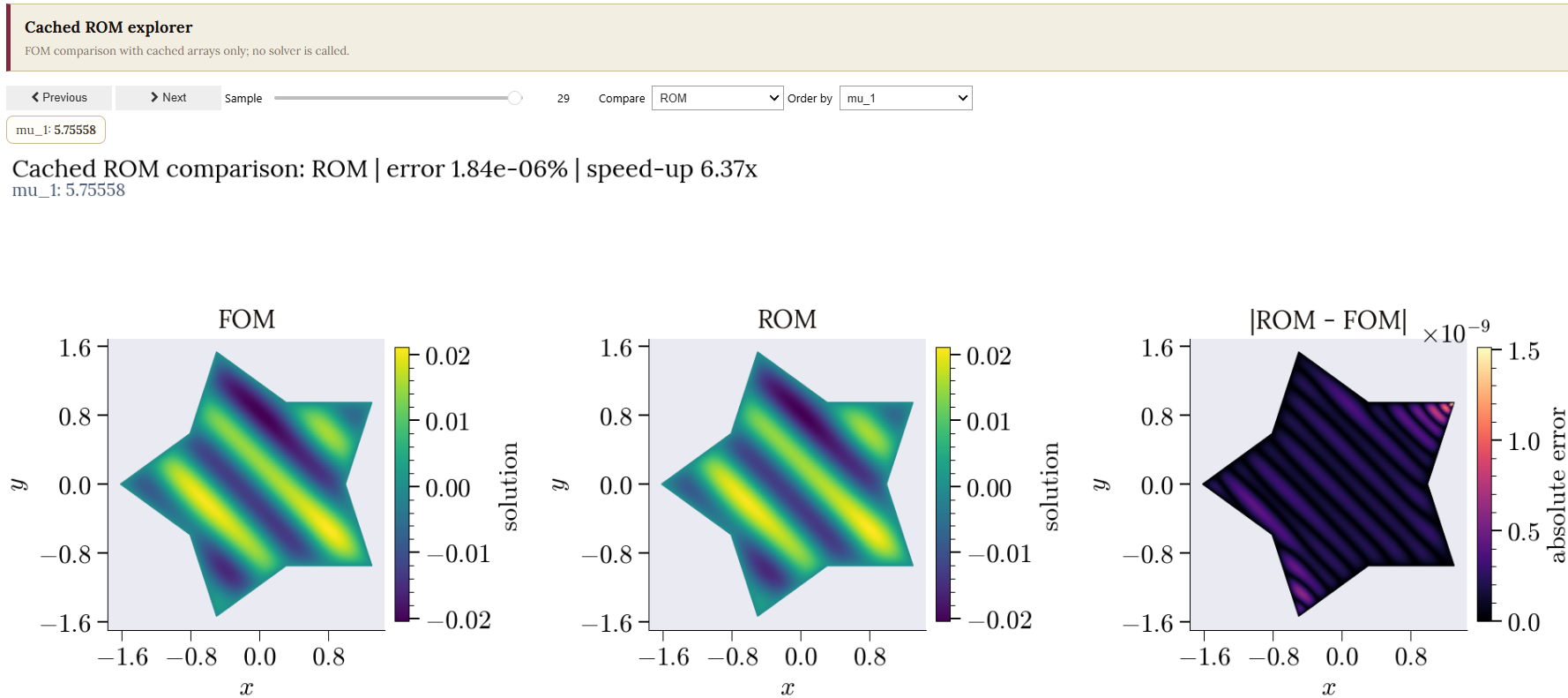}

    \caption{Interactive ROM explorer in \skrom{} for comparing FOM, ROM, and hyper-ROM solutions. The dropdown menu at the top selects the displayed solution type; in this figure, it is set to \textit{ROM}. The explorer also reports the pointwise error, parameter values, relative error, and speed-up.}

    \label{fig:interactive_cached_rom}
\end{figure}

The third tool the \texttt{cached ROM explorer}, depicted in \cref{fig:interactive_cached_rom}, facilitates comparison between a selected full-order solution and a ROM (or hyper-ROM) solution for a specified parameter instance. The tool displays the respective solution profiles, pointwise errors, parameter values, relative error, and speed-up factor. Since these comparison leverages cached arrays, users can seamlessly switch between different parameter samples and reduced models without rerunning the solver.

Collectively, these interactive tools help users visualize,  analyze, and understand basis truncation, POD mode interpretation, ROM accuracy, hyper-ROM comparisons.

\section{Representative model reduction examples with scikit-rom}
\label{sec:examples}

The \skrom{} repository features a diverse collection of examples encompassing heat transfer, computational mechanics, and reaction-diffusion problems. These examples span both static and dynamic scenarios, and include linear as well as nonlinear formulations. The repository demonstrates how to construct and work with full-order, reduced-order, and hyper-reduced models within the framework. Key aspects illustrated include the use of parameterized material properties, explicit domain construction, definition of finite element forms, solver configuration, and systematic model reduction workflows. Together, these examples serve as representative demonstrations for a variety of engineering and physics-based applications.

In this section, we present four representative model order reduction problems chosen from those examples. They cover linear and nonlinear heat transfer and computational mechanics problems in both static and dynamic settings. The results for these four examples were generated using a 12th Gen Intel(R) Core(TM) i7-12700K CPU (32 GB RAM). Each example is presented not only as a computational benchmark but also as a self-contained teaching module that demonstrates a specific concept in projection-based model reduction.

\subsection{Numerical example: one-dimensional nonlinear heat conduction}

\subsubsection{Problem setup and full-order model}

We illustrate the preceding workflow using a one-dimensional nonlinear heat
conduction problem on the interval $[0,L]$, with $L=0.5$~m. This benchmark
problem has been discussed in detail in~\cite{Bhattacharyya2025}, and the
corresponding \skrom{} implementation is available
in~\cite{bhattacharyya_scikitrom_problem1h_github}. The spatial domain is
discretized with linear $P_1$ finite elements on a uniform mesh. A Dirichlet
condition is imposed at the right boundary, with $T(L)=573.15$~K. The bar is
divided at $x=0.4$~m into two material subdomains. Rather than treating the
two regions through separate assembly branches, the implementation defines
Boolean element masks and passes them through the assembly pipeline:

\begin{lstlisting}[language=Python]
break_point = 0.4
bool_arr = element2loc < break_point
regions_mask = {
    'region_1':  np.all(bool_arr, axis=1),
    'region_2': ~np.all(bool_arr, axis=1),
}
global_mask = tuple(regions_mask.items())
\end{lstlisting}

This design allows the same residual and Jacobian routines to handle
heterogeneous material behavior without changing the solver structure. The
conductivity and volumetric source are both nonlinear and parametric. Two
scalar parameters are introduced, $k_p \in [-4,4]$ and $q_p \in
[-1000,1000]$. Here, $u$ denotes the temperature field. In the first
subdomain,
\begin{equation}
  k_1(u;k_p) = 16 + k_p + \frac{2150}{u - 73.15},
  \qquad
  q_1(u;q_p) = q_p + 35000 + \frac{u}{10},
\end{equation}
whereas in the second subdomain,
\begin{equation}
  k_2(u;k_p) = 30 + k_p
             + 2.09\times10^{-2}\,u
             - 1.45\times10^{-5}\,u^2
             + 7.67\times10^{-9}\,u^3,
\end{equation}
\begin{equation}
  q_2(u;q_p) = 10\,q_p + 5000.
\end{equation}
These expressions and their analytical derivatives are implemented in
\texttt{properties.py}.

The full-order model is written in residual form. Let $v$ denote a
finite-element test function that vanishes on the Dirichlet boundary, and let
$\bmu=(k_p,q_p)$ denote the parameter vector. The residual is
\begin{equation}
  \mathcal{R}(u;\bmu)[v]
  =
  \int_0^L
  \left[
      k(u;\bmu)\,
      \frac{\mathrm{d}u}{\mathrm{d}x}
      \frac{\mathrm{d}v}{\mathrm{d}x}
      -
      q(u;\bmu)\,v
  \right]
  \,\mathrm{d}x
  =
  0
  \qquad
  \forall\, v \in V_0 ,
  \label{eq:heat1d_residual}
\end{equation}
where $V_0$ is the homogeneous test space associated with the imposed Dirichlet
condition. The functions $k(u;\bmu)$ and $q(u;\bmu)$ denote the
subdomain-dependent conductivity and source terms defined above.

The corresponding Newton linearization is obtained by differentiating the
residual with respect to $u$. If $\delta u$ denotes the Newton correction, the
Jacobian bilinear form is
\begin{equation}
  \mathcal{J}(u;\bmu)[\delta u,v]
  =
  \int_0^L
  \left[
      k(u;\bmu)\,
      \frac{\mathrm{d}(\delta u)}{\mathrm{d}x}
      \frac{\mathrm{d}v}{\mathrm{d}x}
      +
      \frac{\partial k}{\partial u}(u;\bmu)\,
      \delta u\,
      \frac{\mathrm{d}u}{\mathrm{d}x}
      \frac{\mathrm{d}v}{\mathrm{d}x}
      -
      \frac{\partial q}{\partial u}(u;\bmu)\,
      \delta u\,v
  \right]
  \,\mathrm{d}x .
  \label{eq:heat1d_jacobian}
\end{equation}
Within the problem class, the residual and Jacobian forms are supplied through
\texttt{linear\_forms.py} and \texttt{bilinear\_forms.py}, respectively. The
current Newton iterate and the parameter pair are passed to the assembler
through
\begin{lstlisting}[language=Python]
def assemble_kwargs(self, u, param):
    return dict(u_prev=u, k_param=param[0], q_param=param[1])
\end{lstlisting}
and the subdomain mask is appended inside the assembly routines. The full-order
snapshots are generated by executing the master-class workflow from the problem
notebook:
\begin{lstlisting}[language=Python]
sim = master.fom_simulation(num_snapshots=16)
sim.run_simulation()
\end{lstlisting}
For this example, the workflow produces $16$ training and $16$ test parameter
instances using Sobol sampling over the two-dimensional parameter domain. The
finite-element model uses approximately $1.6\times10^{4}$ elements.

\subsubsection{Projection-based reduced-order model}

With the problem setup and full-order snapshots available, the next step is to
construct a projection-based reduced-order model using the stored training
solutions. In the notebook, this process begins by loading the saved
finite-element metadata and full-order solutions:
\begin{lstlisting}[language=Python]
current_dir = Path().resolve()
rom_dir     = current_dir / "ROM_data"
fos_solutions, sim_data = load_rom_data(
    self=None, rom_data_dir=rom_dir
)
\end{lstlisting}
These files contain the full-order snapshots together with the mesh, basis,
parameter lists, training and testing masks, and other data required for
reduced-model assembly. The training snapshots are then extracted and centered
by subtracting their empirical mean:
\begin{lstlisting}[language=Python]
NLS_train_mean = np.mean(NLS_train, axis=0)
NLS_train_ms   = NLS_train - NLS_train_mean
\end{lstlisting}
Here, \texttt{NLS\_train} stores the nonlinear full-order solution snapshots
used for training, \texttt{NLS\_train\_mean} is their empirical mean, and
\texttt{NLS\_train\_ms} is the corresponding mean-subtracted snapshot matrix.
Centering the data allows the reduced basis to represent fluctuations around a
reference state.

A singular value decomposition is then performed on the centered snapshots, and
the leading left singular vectors define the POD basis~\cite{Bhattacharyya2025}:
\begin{lstlisting}[language=Python]
n_sel, U = svd_mode_selector(
    NLS_train_ms, tolerance=1e-4, modes=True
)
V_sel = U[:, :n_sel]
\end{lstlisting}
In this example, four modes are retained. Let $N$ denote the number of
full-order degrees of freedom and let $r$ denote the number of retained modes.
The reduced basis is denoted by $\mat{U}\in\mathbb{R}^{N\times r}$. In the
code, this basis is stored as \texttt{V\_sel}. For a reduced coordinate vector
$\widehat{\vec{u}}\in\mathbb{R}^{r}$, the reconstructed full-order temperature
vector is written as
\begin{equation}
  \vec{u}
  \approx
  \bar{\vec{u}}_{\mathrm{train}}
  +
  \mat{U}\widehat{\vec{u}},
  \label{eq:ex1_rom_reconstruction}
\end{equation}
where $\bar{\vec{u}}_{\mathrm{train}}\in\mathbb{R}^{N}$ is the empirical mean
of the training snapshots.

After constructing the basis, the notebook initializes the ROM simulation object
by passing the basis, the number of selected modes, and the reference field:
\begin{lstlisting}[language=Python]
rom = master.rom_simulation(
    V_sel=V_sel, n_sel=n_sel, train_ref=NLS_train_mean
)
rom.run_rom_simulation()
\end{lstlisting}
At the implementation level, the nonlinear reduced operators are assembled in
\texttt{problem\_def.py}. For each reduced iterate, the corresponding full-order
field is reconstructed, the full-order Jacobian matrix and residual vector are
assembled, and both are projected onto the reduced space:
\begin{lstlisting}[language=Python]
def reduced_operators(self, u_rom, param):
    u_full = reconstruct_solution(u_rom, self.U, self.T_ref)
    J_mat  = self.fom_operators(u_full, param)
    R_vec  = self.fom_rhs(u_full, param)
    J_red  = self.U.T @ (J_mat @ self.U)
    R_red  = self.U.T @ R_vec
    return J_red, R_red
\end{lstlisting}
Mathematically, this corresponds to
\begin{equation}
  \widehat{\mat{J}}(\widehat{\vec{u}};\bmu)
  =
  \mat{U}^{\top}
  \mat{J}(\vec{u};\bmu)
  \mat{U},
  \qquad
  \widehat{\vec{R}}(\widehat{\vec{u}};\bmu)
  =
  \mat{U}^{\top}
  \vec{R}(\vec{u};\bmu),
  \label{eq:ex1_reduced_newton}
\end{equation}
where $\vec{u}$ is reconstructed from \eqref{eq:ex1_rom_reconstruction}.
Here, $\widehat{\mat{J}}\in\mathbb{R}^{r\times r}$ and
$\widehat{\vec{R}}\in\mathbb{R}^{r}$ define the reduced Newton system.
Although the Newton solve is reduced to dimension $r$, the standard ROM still
assembles full-order matrices and residual vectors during the online stage,
which limits the speed-up for this nonlinear example. After solving, the
reduced-order outputs can be accessed as follows:
\begin{lstlisting}[language=Python]
NLS_rom           = np.asarray(rom.rom_solutions)
ROM_speed_up      = rom.speed_up
ROM_relative_error = rom.rom_error
\end{lstlisting}

Here, \texttt{NLS\_rom} stores the reconstructed ROM solutions, while
\texttt{ROM\_speed\_up} and \texttt{ROM\_relative\_error} store the timing and
accuracy measures over the test set.

\begin{figure}[t]
  \centering
  \begin{subfigure}{0.45\linewidth}
    \centering
    \includegraphics[width=\linewidth]{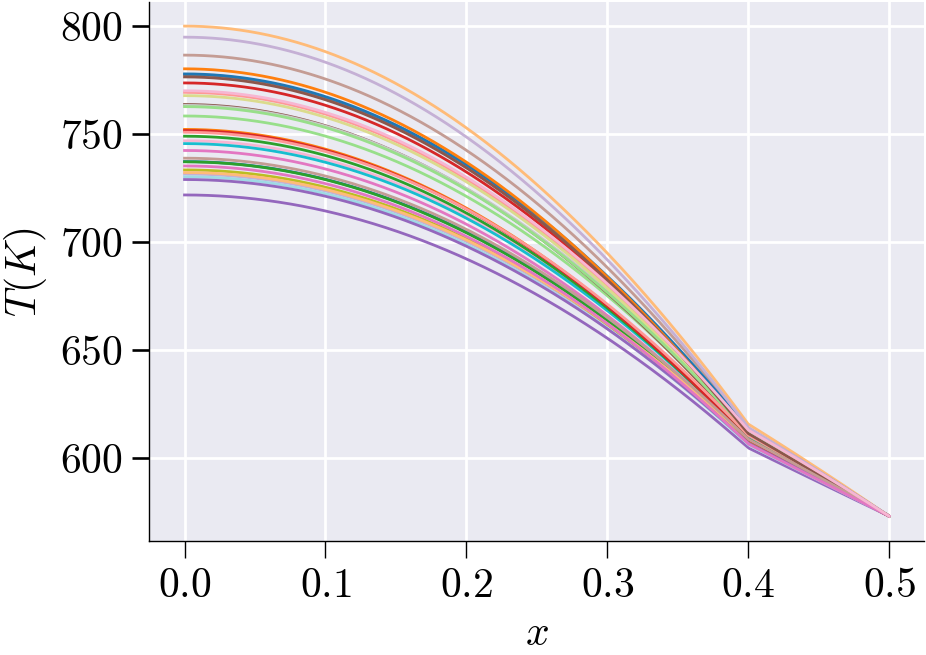}
    \caption{}
    \label{fig:ex1_fom}
  \end{subfigure}
  \hfill
  \begin{subfigure}{0.45\linewidth}
    \centering
    \includegraphics[width=\linewidth]{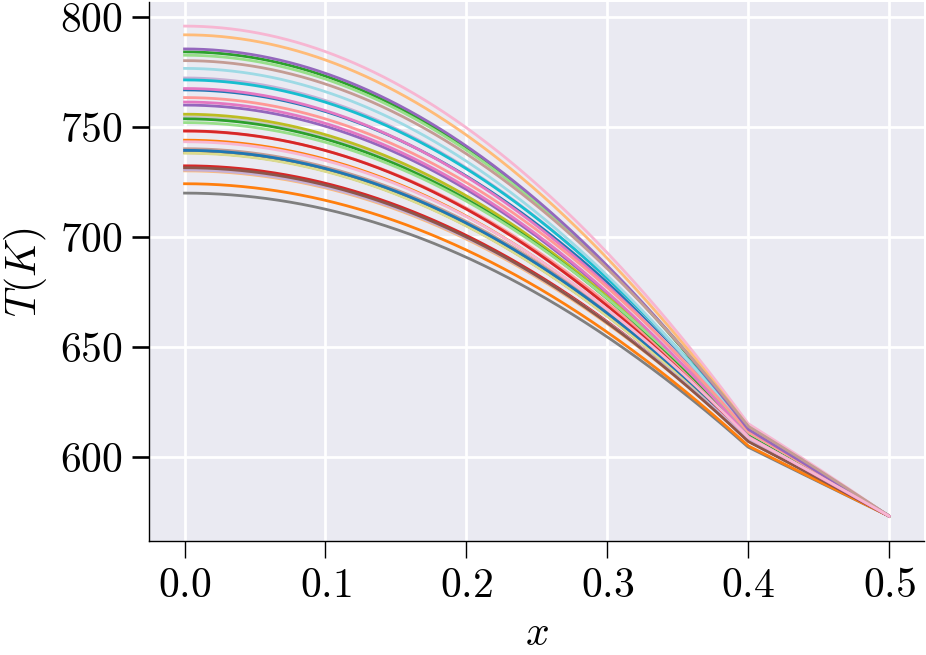}
    \caption{}
    \label{fig:ex1_rom}
  \end{subfigure}
  \begin{subfigure}{0.7\linewidth}
    \centering
    \includegraphics[width=\linewidth]{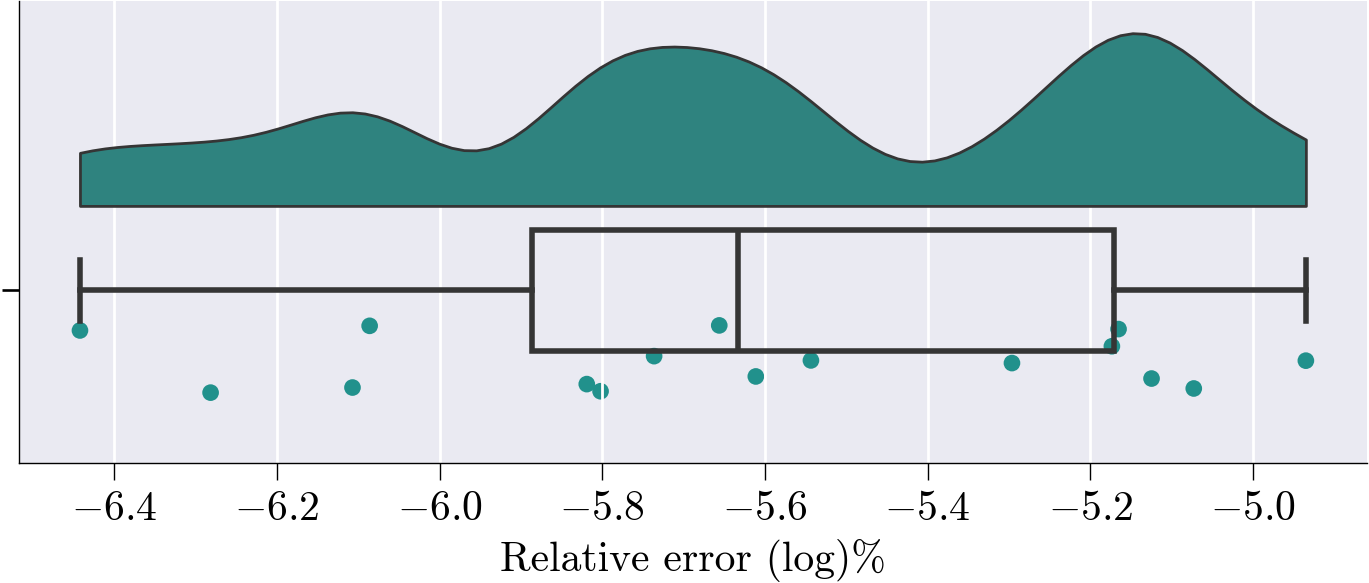}
    \caption{}
    \label{fig:ex1_rom_error}
  \end{subfigure}
  \begin{subfigure}{0.7\linewidth}
    \centering
    \includegraphics[width=\linewidth]{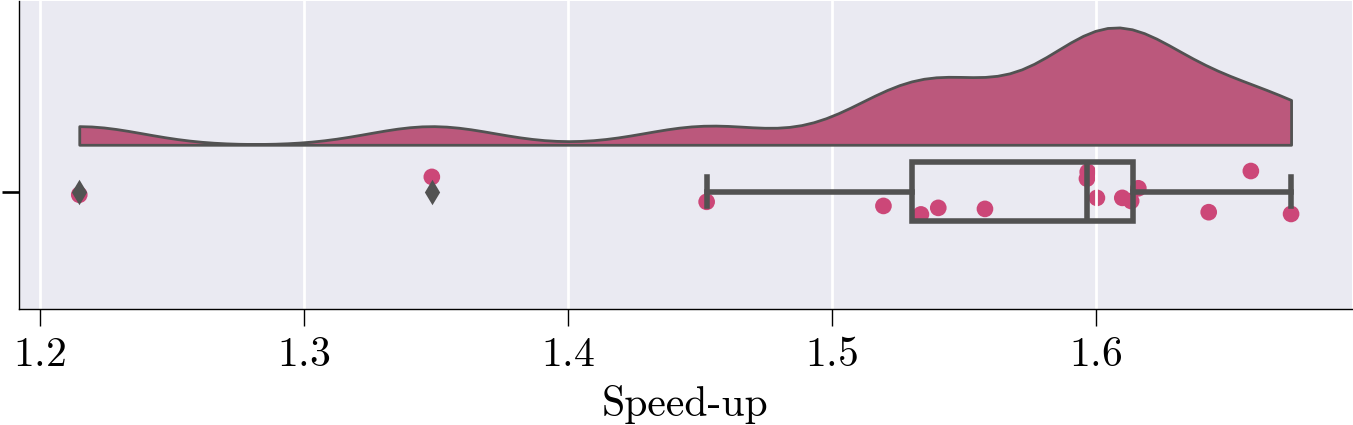}
    \caption{}
    \label{fig:ex1_rom_speedup}
  \end{subfigure}
  \caption{Comparison of the full-order model and the four-dimensional ROM for Example~1. 
  (a) FOM temperature solutions. 
  (b) ROM temperature solutions. 
  (c) Relative ROM error over the test set. 
  (d) ROM speed-up over the corresponding full-order solves.}
  \label{fig:ex1_combined}
\end{figure}

The reduced order model (ROM) results are compared with the full order model (FOM) results in \cref{fig:ex1_combined}. 
The four-dimensional ROM reproduces the temperature profiles with very small relative error, but the speed-up remains limited because the nonlinear residual and Jacobian are still assembled in the full-order space during the online stage.

\subsubsection{ECSW, ECM, and DEIM hyper-reduction}

While the projection-based ROM reduces the size of the nonlinear Newton system,
it still requires full-order operator assembly during the online phase. This
explains the limited ROM speed-up observed in \cref{fig:ex1_rom_speedup}.
Hyper-reduction is introduced to reduce this remaining cost. In \skrom{}, this
is done without redefining the underlying full-order problem: the same problem
definition and reduced basis are used, but the online operator evaluation is
replaced by a sparse approximation.

For ECSW, the notebook first constructs a reduced residual operator and
collects residual contribution snapshots across the training set:
\begin{lstlisting}[language=Python]
from linear_forms import R
training_params = param_list[train_mask]

Residual = LinearFormROM(
    R, basis, V_sel, free_dofs=sim_data['free_dofs']
)
q_mus = collect_residuals(
    NLS_train_ms, NLS_train_mean,
    V_sel, reconstruct_solution,
    Residual, training_params,
    prob.assemble_kwargs,
    extra_kwargs=dict(global_mask=sim_data['global_mask'])
)
\end{lstlisting}
The matrix \texttt{q\_mus} collects projected residual contributions over the
training parameters and snapshots. These data are then compressed, and the ECSW
weights are computed by solving a sparse nonnegative approximation problem:
\begin{lstlisting}[language=Python]
Uh, s, Vh = np.linalg.svd(q_mus, full_matrices=False)
V_q_eff   = Vh[:20]

x_nnls_m, err = hyperreduce(
    V_q_eff, verbosity=2,
    const_tol=1e-8, zero_tol=1e-8, plot=False
)
\end{lstlisting}
The vector \texttt{x\_nnls\_m} contains the ECSW element weights. Its nonzero
entries identify the elements retained in the hyper-reduced mesh, and the
corresponding values define their weights in the sparse element summation. The
ECSW simulation is then executed as follows:
\begin{lstlisting}[language=Python]
rom.run_hyper_rom_simulation_ecsw(x_nnls_m)
\end{lstlisting}

\begin{figure}[t]
  \centering
  \begin{subfigure}{0.7\linewidth}
    \centering
    \includegraphics[width=\linewidth]{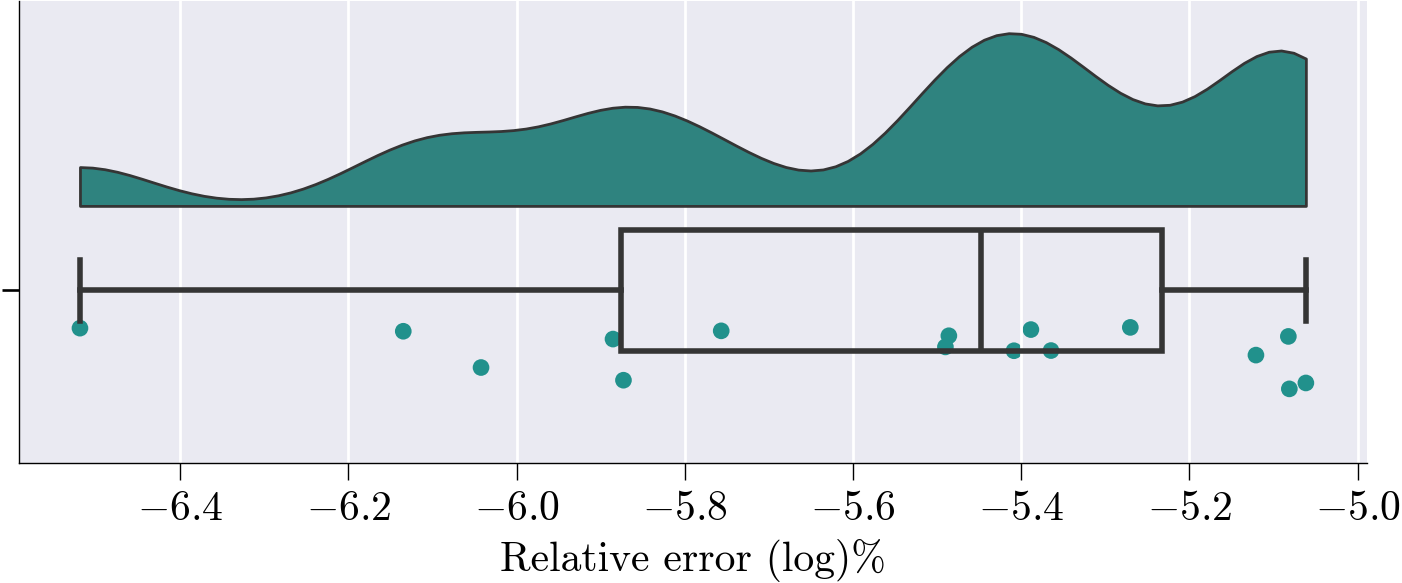}
    \caption{}
  \end{subfigure}
  \begin{subfigure}{0.7\linewidth}
    \centering
    \includegraphics[width=\linewidth]{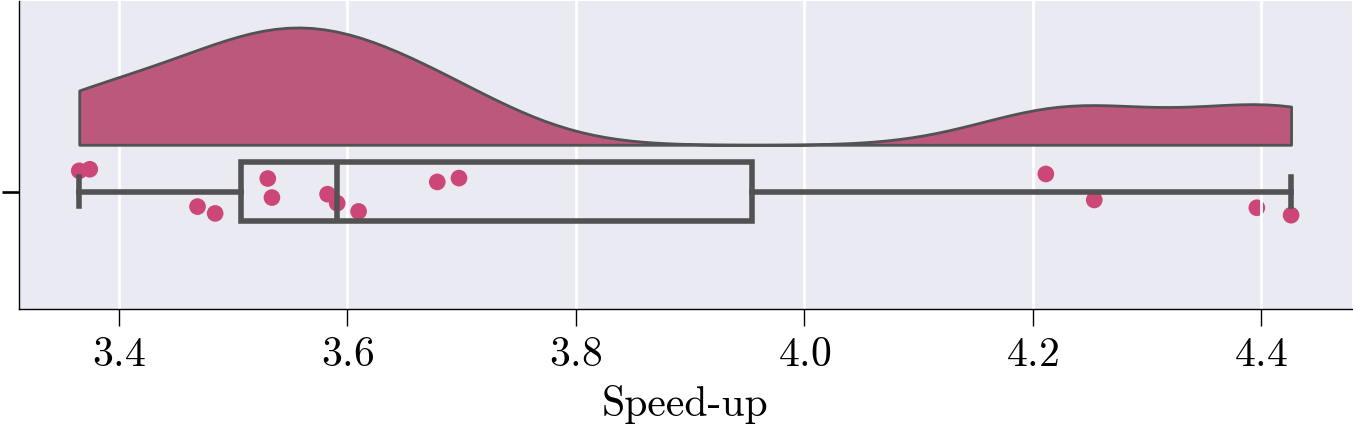}
    \caption{}
  \end{subfigure}
\caption{ECSW-based HyperROM performance for Example~1. 
(a) Relative error with respect to the FOM solution. 
(b) Speed-up obtained by the ECSW HyperROM over the corresponding full-order solves.}
  \label{fig:hrom_ecsw}
\end{figure}

\begin{figure}[t]
  \centering
  \begin{subfigure}{0.7\linewidth}
    \centering
    \includegraphics[width=\linewidth]{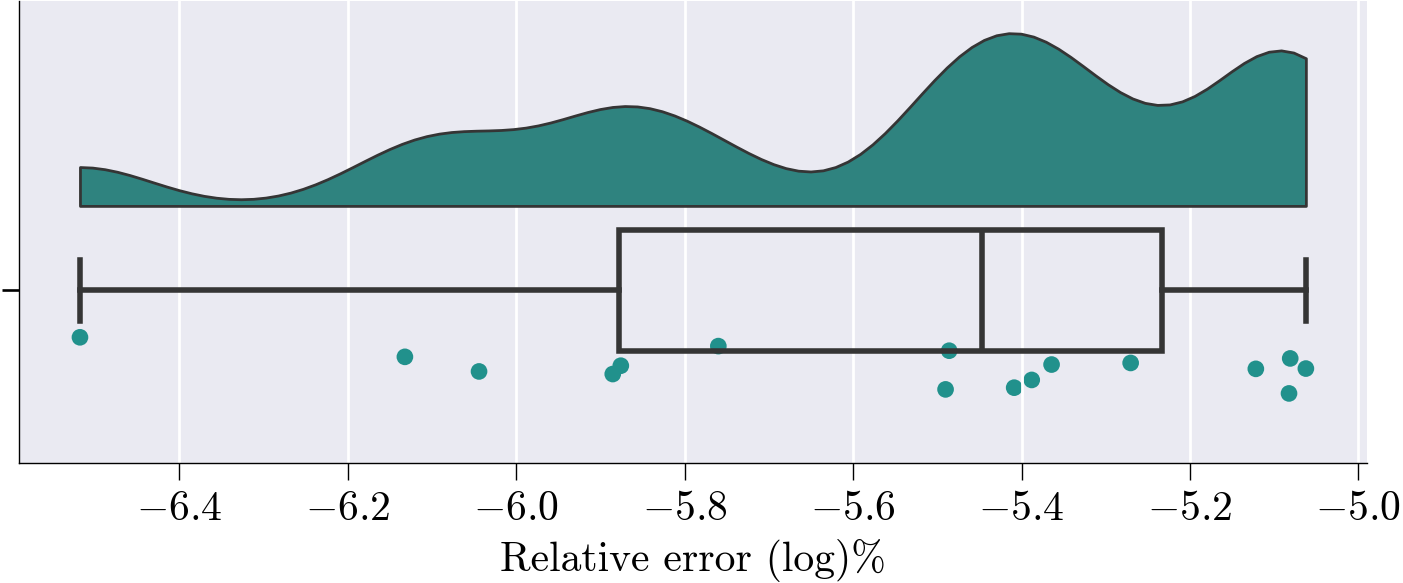}
    \caption{}
  \end{subfigure}
  \begin{subfigure}{0.7\linewidth}
    \centering
    \includegraphics[width=\linewidth]{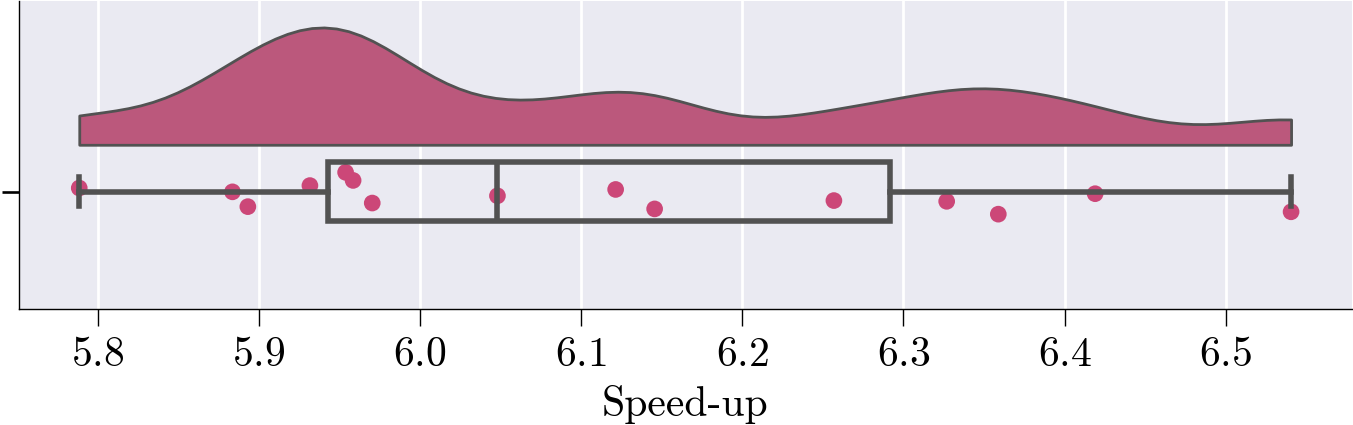}
    \caption{}
  \end{subfigure}
\caption{ECM-based HyperROM performance for Example~1. 
(a) Relative error with respect to the FOM solution. 
(b) Speed-up obtained by the ECM HyperROM over the corresponding full-order solves.}
  \label{fig:hrom_ecm}
\end{figure}

The ROM accuracy and speed-up obtained with ECSW are shown in
\cref{fig:hrom_ecsw}. For ECM, a similar workflow is used, but the sparse
approximation is constructed at the quadrature-point level rather than the
element level:
\begin{lstlisting}[language=Python]
ecm_data = train_ecm.ECM_from_skrom(
    NLS_train_ms         = NLS_train_ms,
    NLS_train_mean       = NLS_train_mean,
    V_sel                = V_sel,
    reconstruct_solution = reconstruct_solution,
    Residual             = Residual,
    training_params      = training_params,
    assemble_kwargs      = prob.assemble_kwargs,
    basis                = basis,
    extra_kwargs         = dict(global_mask=sim_data["global_mask"]),
    tol                  = 1e-8,
    plot                 = False,
    svd_rank             = 30,
)
gauss_weight_ecm = ecm_data["gauss_weight_ecm"]
\end{lstlisting}
The array \texttt{gauss\_weight\_ecm} stores the ECM cubature weights over
element--Gauss-point pairs. Nonzero entries indicate the integration points
retained during the online assembly. These weights are then passed to the ECM
solver:
\begin{lstlisting}[language=Python]
rom.run_hyper_rom_simulation_ecm(z=gauss_weight_ecm)
\end{lstlisting}
Thus, ECSW and ECM follow the same sparse-weighting principle, but ECSW selects
weighted elements whereas ECM selects weighted quadrature points. The ROM
accuracy and speed-up obtained with ECM are shown in \cref{fig:hrom_ecm}.

The DEIM approach uses snapshots of the nonlinear term itself rather than
residual contribution snapshots:
\begin{lstlisting}[language=Python]
prob.basis       = basis
prob.global_mask = sim_data['global_mask']
prob.mesh        = sim_data['mesh'].item()

F_nl = compute_nonlinear_snapshots(
    non_linear_func = prob.f_nl,
    fos_solutions   = fos_solutions[train_mask],
    param_list      = param_list[train_mask]
)
\end{lstlisting}
Here, \texttt{F\_nl} is the nonlinear snapshot matrix. Its columns store
nonlinear term evaluations at the training solutions and parameter values. From
these snapshots, the DEIM interpolation structure is built:
\begin{lstlisting}[language=Python]
deim_class = deim_module.deim(
    prob.mesh, F_nl, V_sel,
    tol_f=1e-7, extra_modes=0
)
deim_mat, sampled_rows = deim_class.select_elems()
xi_deim = deim_class.xi
\end{lstlisting}
The variable \texttt{sampled\_rows} contains the selected interpolation degrees
of freedom, \texttt{deim\_mat} is the interpolation matrix used to reconstruct
the reduced nonlinear contribution, and \texttt{xi\_deim} marks the elements
associated with the selected degrees of freedom. The DEIM-based online
simulation is then launched as
\begin{lstlisting}[language=Python]
rom.run_hyper_rom_simulation_deim(
    z            = xi_deim,
    deim_mat     = deim_mat,
    sampled_rows = sampled_rows,
)
\end{lstlisting}
Here, the nonlinear term is approximated through interpolation rather than
weighted summation, while the reduced basis and the full problem definition
remain unchanged.

\begin{figure}[t]
  \centering
  \begin{subfigure}{0.7\linewidth}
    \centering
    \includegraphics[width=\linewidth]{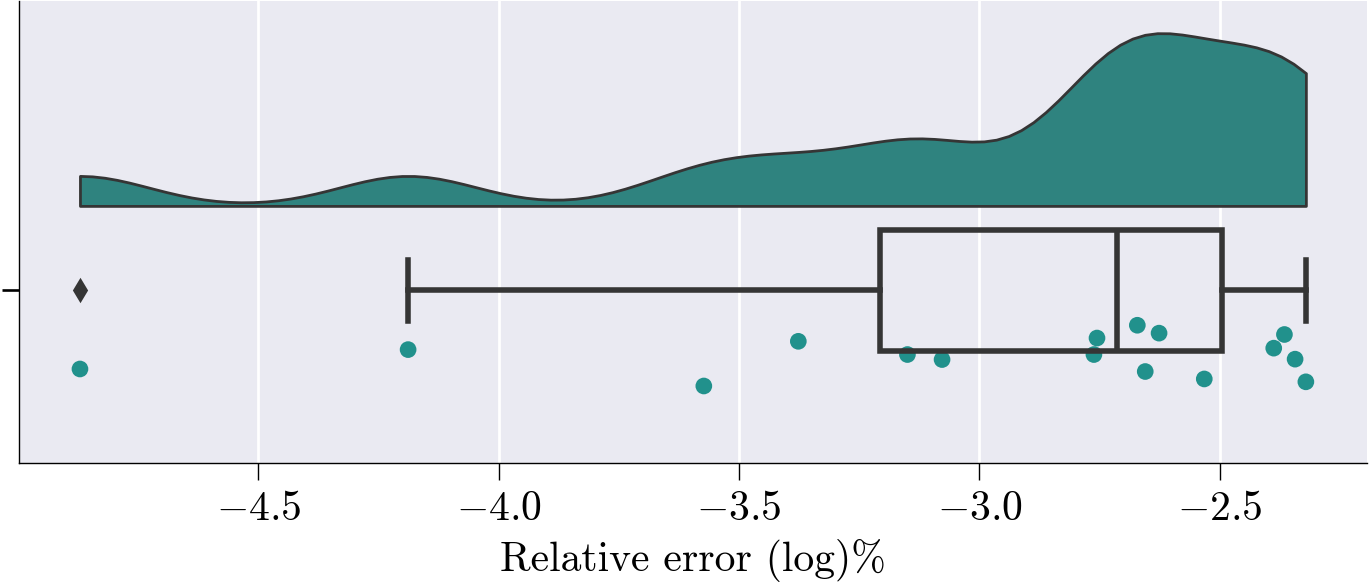}
    \caption{}
  \end{subfigure}
  \begin{subfigure}{0.7\linewidth}
    \centering
    \includegraphics[width=\linewidth]{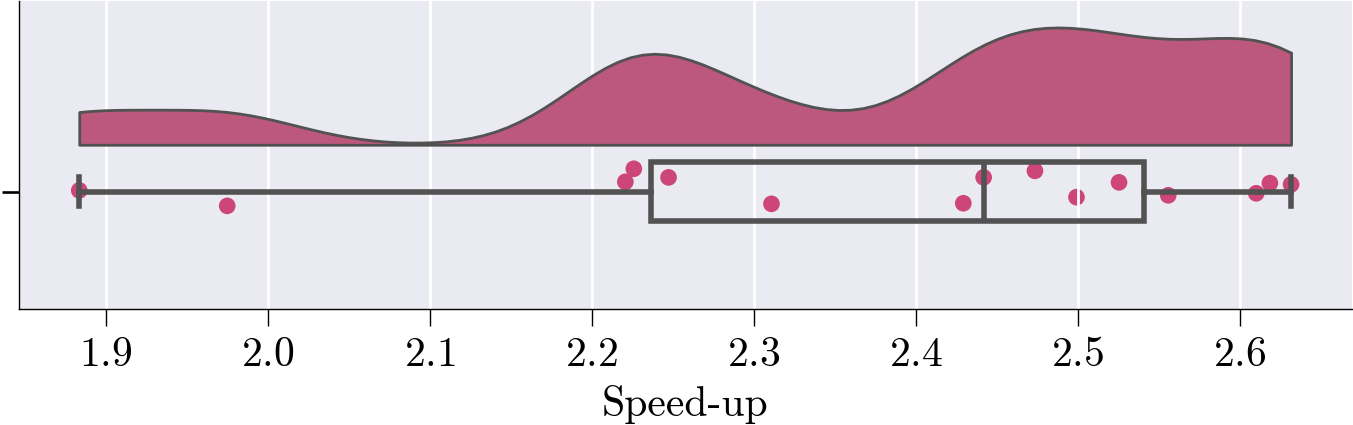}
    \caption{}
  \end{subfigure}
\caption{DEIM-based HyperROM performance for Example~1. 
(a) Relative error with respect to the FOM solution. 
(b) Speed-up obtained by the DEIM HyperROM over the corresponding full-order solves.}
  \label{fig:hrom_deim}
\end{figure}

\subsection{Numerical example: two-dimensional transient linear heat conduction}

\subsubsection{Problem setup and full-order model}

As a second benchmark, we consider a two-dimensional transient heat conduction
problem on the unit square $\Omega=[0,1]^2$~\cite{bhattacharyya_scikitrom_transient_heat_2d}.
Let $T(\vec{x},t;\bmu)$ denote the temperature field, where $\vec{x}=(x,y)$
and $\bmu$ is the parameter vector. The governing equation is
\begin{equation}
  \frac{\partial T}{\partial t}
  - \nabla \cdot \bigl(k\,\nabla T\bigr)
  = q\,g(\vec{x})
  \qquad \text{in } \Omega \times (0,t_f],
\end{equation}
where $k$ is the conductivity scaling and $q$ is the source-amplitude scaling.
The parameter vector is
\begin{equation}
  \bmu = [k,\;q]^{\top},
\end{equation}
with
\begin{equation}
  k \in [0.1,5.0], \qquad q \in [0,5.0].
\end{equation}
The source term has a fixed Gaussian shape centered at $(0.5,0.5)$ with width
$\sigma=0.12$:
\begin{equation}
  g(x,y) = \exp\!\left(
    -\frac{(x-0.5)^2+(y-0.5)^2}{2(0.12)^2}
  \right).
\end{equation}
The spatial discretization uses linear triangular elements on a structured
$25\times25$ mesh of the unit square, giving $1250$ triangular elements.
Homogeneous Dirichlet boundary conditions are imposed on the full outer
boundary, and the initial condition is
\begin{equation}
  T(\vec{x},0;\bmu) = 0.
\end{equation}
Time integration is performed with implicit Euler using
\begin{equation}
  \Delta t = 0.02, \qquad t_f = 0.4,
\end{equation}
which gives $21$ time levels including the initial state.

After spatial discretization, let $\mathbf{T}(t;\bmu)\in\mathbb{R}^{N}$ denote
the vector of nodal temperature unknowns, where $N$ is the number of
unconstrained finite-element degrees of freedom. The semi-discrete system is
\begin{equation}
  \mat{M}\dot{\mathbf{T}}(t;\bmu)
  + \mat{K}(\bmu)\mathbf{T}(t;\bmu)
  = \mathbf{f}(\bmu),
\end{equation}
where $\mat{M}$ is the mass matrix, $\mat{K}(\bmu)$ is the conductivity
matrix, and $\mathbf{f}(\bmu)$ is the load vector associated with the Gaussian
source. Since the problem is affine in the parameters,
\begin{equation}
  \mat{K}(\bmu) = k\,\mat{K}_0,
  \qquad
  \mathbf{f}(\bmu) = q\,\mathbf{f}_0,
\end{equation}
where $\mat{K}_0$ and $\mathbf{f}_0$ are the parameter-independent reference
stiffness matrix and load vector.

A reduced basis consisting of three left singular vectors is constructed from
the full-order snapshots over the training set. These modes capture $99\%$ of
the data variance. Because this example is linear and affine, standard Galerkin
projection is sufficient and no hyper-reduction is introduced. With
$\mat{U}\in\mathbb{R}^{N\times r}$ denoting the reduced basis, the reduced
operators are
\begin{equation}
  \widehat{\mat{M}} = \mat{U}^{\top}\mat{M}\mat{U},
  \qquad
  \widehat{\mat{K}}(\bmu) = k\,\widehat{\mat{K}}_0,
  \qquad
  \widehat{\mathbf{f}}(\bmu) = q\,\widehat{\mathbf{f}}_0,
\end{equation}
where $\widehat{\mat{M}}\in\mathbb{R}^{r\times r}$,
$\widehat{\mat{K}}_0\in\mathbb{R}^{r\times r}$, and
$\widehat{\mathbf{f}}_0\in\mathbb{R}^{r}$ are the projected mass, stiffness,
and load quantities.

\subsubsection{ROM results}

\begin{figure}
  \centering
  \includegraphics[width=\linewidth]{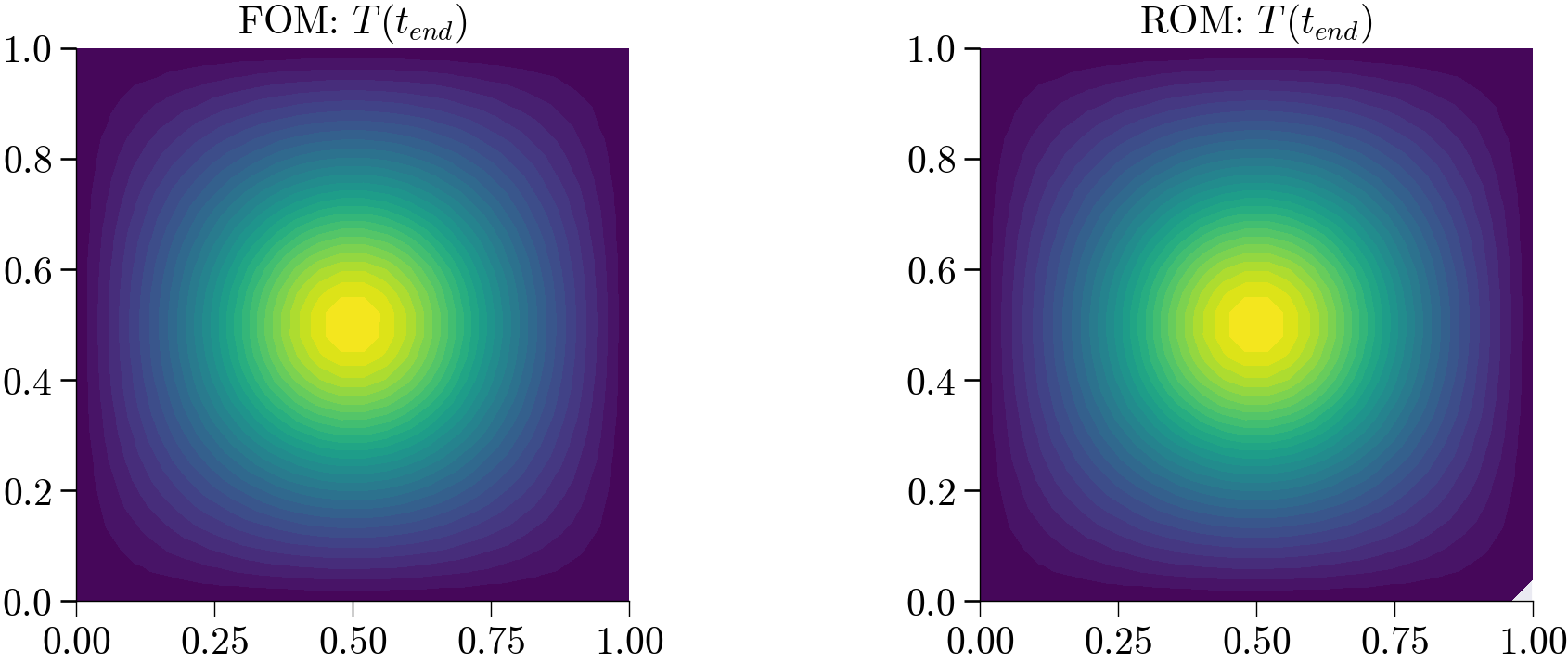}
\caption{Representative temperature fields from the full-order model and the ROM at the final time instant for the two-dimensional transient heat conduction problem in Example~2.}
  \label{fig:heat2d_fields}
\end{figure}

\begin{figure}
  \centering
  \includegraphics[width=0.7\linewidth]{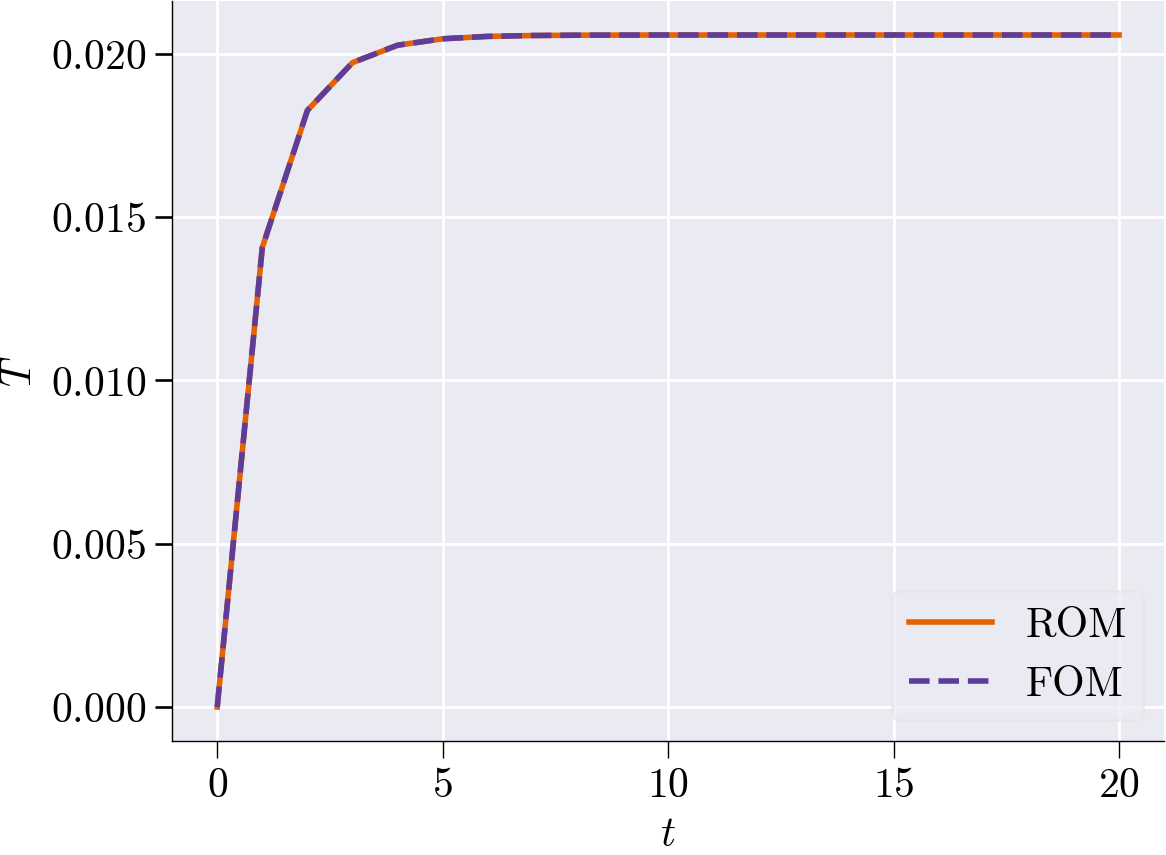}
\caption{Temperature evolution at node~300 for the two-dimensional transient heat conduction problem in Example~2. The curves compare the full-order model and ROM predictions over the simulation time interval.}
  \label{fig:heat2d_history}
\end{figure}

\begin{table}[htbp]
  \centering
  \caption{Quantitative summary of ROM performance for the two-dimensional heat
    transfer problem in Ex.~2.}
  \label{tab:ex2_rom}
  \begin{tabular}{lc}
    \hline
    \textbf{Metric} & \textbf{Value} \\
    \hline
    Average speed-up                             & $15$                       \\
    Relative $L_2$ error over all parameters     & $1.5582\times10^{-4}$      \\
    Relative $L_\infty$ error over all parameters & $3.8189\times10^{-3}$     \\
    RMSE                                         & $2.5804\times10^{-6}$      \\
    $R^2$ score                                  & $1.0000$                   \\
    \hline
  \end{tabular}
\end{table}

Figure~\ref{fig:heat2d_fields} shows representative temperature fields from the
full-order model and the ROM at the final time instant. The reduced solution
reproduces the transient diffusion pattern generated by the centered Gaussian
source over the parameter range considered. \Cref{fig:heat2d_history} shows the
temperature history at node~300 for both FOM and ROM. The ROM follows the
full-order response closely throughout the simulation window. A summary of the
ROM performance is reported in \cref{tab:ex2_rom}.

This example demonstrates the standard projection-based workflow of \skrom{}
on a transient linear heat conduction problem. The full-order model generates
solution snapshots over the two-parameter space $(k,q)$, the reduced basis is
extracted from these data, and the online stage proceeds through affine reduced
operators.

\subsection{Numerical example: static nonlinear hyperelasticity}

\subsubsection{Problem setup and full-order model}

As a third benchmark, we consider the \skrom{} implementation of a static
finite-strain hyperelasticity problem, following the \texttt{scikit-fem}
hyperelasticity example~\cite{skfem2020,skfem_ex43} and as described
in~\cite{bhattacharyya_scikitrom_static_nonlinear_problem1h_github}. The
reference domain is the block
\begin{equation}
  \Omega_0 = [0,10]\times[0,1]\times[0,1],
\end{equation}
discretized with trilinear hexahedral elements on a structured mesh of
$50\times5\times5$ cells, giving $1250$ elements. The displacement field is
denoted by $\vec{u}(\vec{X};\bmu)$, where $\vec{X}\in\Omega_0$ is the
reference coordinate and $\bmu$ is the material-parameter vector. The
displacement is approximated using a vector-valued first-order finite element
basis. The material follows a compressible neo-Hookean law.

The deformation gradient and its determinant are
\begin{equation}
  \mat{F} = \mat{I} + \nabla_{\!X}\vec{u},
  \qquad
  J = \det\mat{F},
\end{equation}
where $\nabla_{\!X}$ denotes the gradient with respect to the reference
coordinates. The Lam\'e parameters are treated as input parameters,
\begin{equation}
  \bmu = [\mu,\lambda]^{\top},
  \qquad
  \mu \in [0.5,1],\quad \lambda \in [1,2].
\end{equation}
The strain-energy density per unit reference volume is
\begin{equation}
  \psi(\mat{F})
  = \frac{\mu}{2}(I_1 - 3) - \mu\ln J
  + \frac{\lambda}{2}(\ln J)^2,
\end{equation}
where
\begin{equation}
  I_1 = \mathrm{tr}(\mat{F}^{T}\mat{F}) = \mat{F}:\mat{F}
\end{equation}
is the first invariant of the right Cauchy--Green deformation tensor.

The weak form seeks $\vec{u}\in\mathcal{V}$ such that
\begin{equation}
  \mathcal{R}(\vec{u};\bmu;\vec{v}) = 0
  \qquad \forall\,\vec{v}\in\mathcal{V}_0,
\end{equation}
where $\mathcal{V}$ is the trial space satisfying the prescribed displacement
boundary conditions and $\mathcal{V}_0$ is the corresponding homogeneous test
space. The residual is
\begin{equation}
  \mathcal{R}(\vec{u};\bmu;\vec{v})
  = \int_{\Omega_0} \mat{P}:\nabla_{\!X}\vec{v}\,\mathrm{d}\Omega,
\end{equation}
with first Piola--Kirchhoff stress
\begin{equation}
  \mat{P} = \mu\,\mat{F} + (\lambda\ln J - \mu)\mat{F}^{-T}.
\end{equation}
The consistent tangent is used in Newton's method. If $\Delta\vec{u}$ denotes
the Newton correction and $\vec{v}$ denotes the virtual displacement, the
tangent form is
\begin{align}
  \Delta\mathcal{R}(\vec{u};\Delta\vec{u},\vec{v})
  = \int_{\Omega_0} \bigg[
    &\mu\,\nabla_{\!X}\Delta\vec{u}:\nabla_{\!X}\vec{v}
    - (\lambda\ln J - \mu)\,
      \mathrm{tr}\!\left(
        \nabla_{\!X}\Delta\vec{u}\,\mat{F}^{-1}
        \nabla_{\!X}\vec{v}\,\mat{F}^{-1}
      \right)
    \nonumber\\
    &+ \lambda\,
      \mathrm{tr}\!\left(\nabla_{\!X}\vec{v}\,\mat{F}^{-1}\right)
      \mathrm{tr}\!\left(\nabla_{\!X}\Delta\vec{u}\,\mat{F}^{-1}\right)
  \bigg]\,\mathrm{d}\Omega.
\end{align}

Boundary conditions match the reference problem: the left face is fixed and the
right face is prescribed with incremental Dirichlet displacements. For each
load factor
\begin{equation}
  c = \frac{s+1}{n_{\mathrm{steps}}},
\end{equation}
where $s$ is the load-step index, the right face is axially shortened and
rotated about the $x$-axis. The prescribed right-face displacement is
\begin{align}
  u_1 &= -0.1\,c, \\
  u_2 &= y\cos(c\pi) - z\sin(c\pi) - y, \\
  u_3 &= y\sin(c\pi) + z\cos(c\pi) - z.
\end{align}
At the final step, the imposed axial displacement is $-0.1$ and the imposed
rotation is $\pi$. Newton's method is used with up to $10$ iterations per load
step and a convergence tolerance of $10^{-3}$ on the displacement-increment
norm.

In \skrom{}, the residual and tangent are defined in \texttt{linear\_forms.py}
and \texttt{bilinear\_forms.py}, respectively, while the deformation gradient
is defined in \texttt{properties.py}. The current displacement field and
parameter pair are passed through \texttt{problem\_def.py}:
\begin{lstlisting}[language=Python]
def assemble_kwargs(self, uv, param):
    return dict(displacement=uv, mu=param[0], lmbda=param[1])
\end{lstlisting}
This keeps the constitutive evaluation parameterized by $\mu$ and $\lambda$
while preserving the same assembly structure for all sampled parameter values.

Training and testing samples are drawn in the $(\mu,\lambda)$ parameter space
by Sobol sampling, with $32$ training samples and $32$ testing samples.
Full-order displacement snapshots are used to construct the reduced basis. The
reduced model projects the nonlinear residual and tangent onto this basis:
\begin{equation}
  \widehat{\mathbf{R}}(\widehat{\mathbf{u}};\bmu)
  = \mat{U}^{\top}\mathbf{R}
    \!\left(\mat{U}\widehat{\mathbf{u}}
           +\mathbf{u}_{\mathrm{ref}};\bmu\right),
\end{equation}
where $\widehat{\mathbf{u}}$ is the reduced displacement coordinate vector,
$\mat{U}$ is the displacement basis, and $\mathbf{u}_{\mathrm{ref}}$ is the
reference displacement field used in the ROM reconstruction. The reduced
tangent is constructed analogously. Alongside the standard ROM, an ECSW
hyper-reduction is included, in which residuals and tangents are assembled using
precomputed sparse element weights through
\texttt{LinearFormHYPERROM\_ecsw} and \texttt{BilinearFormHYPERROM\_ecsw}.

\subsubsection{ROM and HyperROM results}

\begin{figure}[t]
  \centering
  \includegraphics[width=\textwidth]{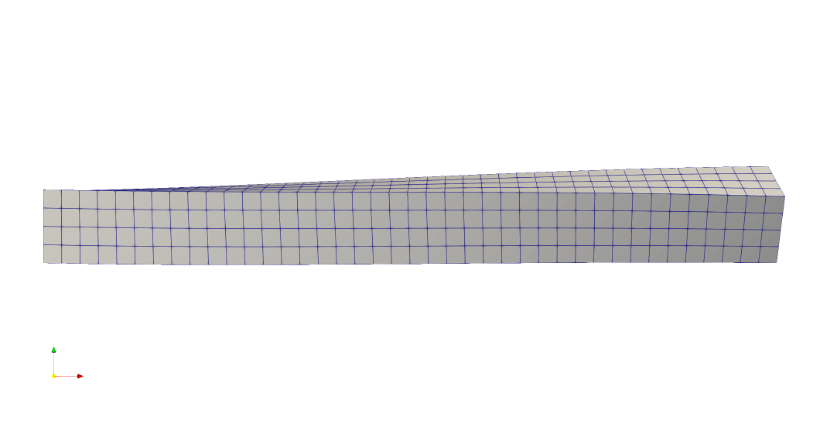}
\caption{Representative deformed configurations for the static nonlinear hyperelasticity problem in Example~3. The figure compares the full-order model, the standard ROM, and the ECSW HyperROM for selected parameter instances.}
  \label{fig:hyperelastic_fields}
\end{figure}

Figure~\ref{fig:hyperelastic_fields} compares representative deformed
configurations obtained from the full-order model, the standard ROM, and the
ECSW HyperROM for selected parameter instances. These results show that the
reduced models retain the dominant deformation pattern under the combined
compression--torsion loading.

\begin{figure}[t]
  \centering
  \begin{subfigure}[b]{\linewidth}
    \centering
    \includegraphics[width=\linewidth]{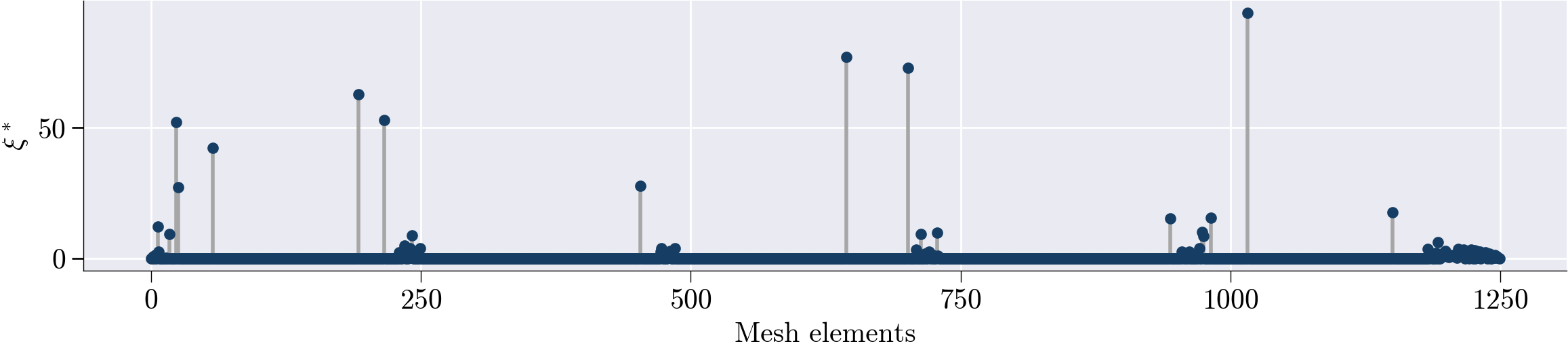}
    \caption{}
    \label{fig:he_a}
  \end{subfigure}
  \begin{subfigure}[b]{0.7\linewidth}
    \centering
    \includegraphics[width=\linewidth]{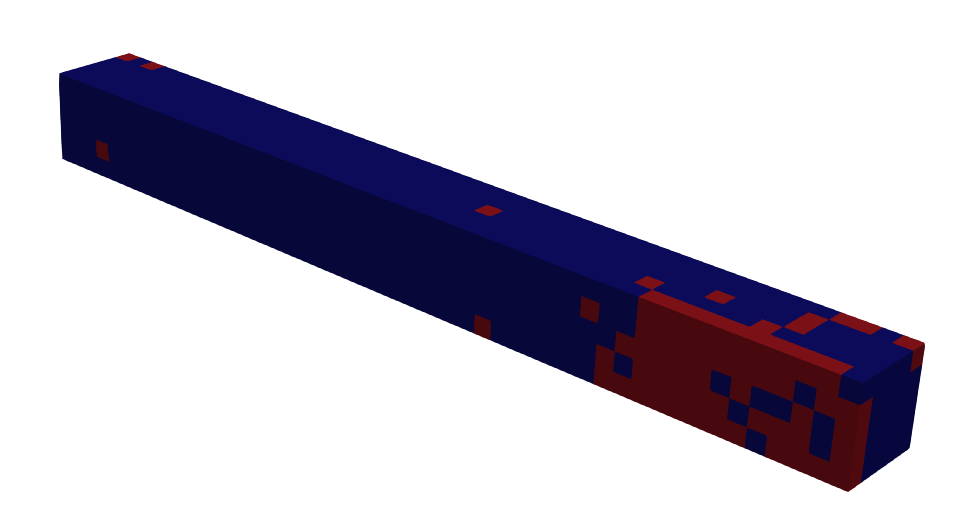}
    \caption{}
    \label{fig:he_b}
  \end{subfigure}
\caption{ECSW sampling pattern for the static nonlinear hyperelasticity problem in Example~3. 
(a) Nonzero ECSW element weights plotted against element index. 
(b) Spatial locations of the selected elements in the three-dimensional domain.}
  \label{fig:hyperelastic_ecsw}
\end{figure}

Since ECSW is implemented for this example, the sparse integration pattern can
also be visualized. Figure~\ref{fig:he_a} shows the selected elements and their
associated weights in the hyper-reduced model. \Cref{fig:he_b} shows the
locations of the selected elements over the three-dimensional domain. A
quantitative summary of the ROM and HyperROM performance is reported in
\cref{tab:ex3_hyperelastic_results}.

\begin{table}[htbp]
  \centering
  \caption{Comparison of ROM and ECSW hyper-reduced ROM accuracy for the static
    nonlinear hyperelasticity problem in Ex.~3.}
  \label{tab:ex3_hyperelastic_results}
  \begin{tabular}{lcc}
    \hline\hline
    Metric                     & ROM                    & ECSW HyperROM        \\
    \hline
    Average speed-up           & $1$                    & $6$                   \\
    Relative $L_2$ error       & $2.2812\times10^{-9}$  & $1.0801\times10^{-2}$ \\
    Relative $L_\infty$ error  & $8.1744\times10^{-9}$  & $6.8066\times10^{-3}$ \\
    RMSE                       & $3.0301\times10^{-10}$ & $1.4348\times10^{-3}$ \\
    $R^2$ score                & $1.0000$               & $0.9999$              \\
    \hline
  \end{tabular}
\end{table}

This example demonstrates the nonlinear solid-mechanics workflow of \skrom{}
on a finite-strain hyperelastic problem. The full-order model is solved over a
two-parameter material space, a reduced basis is constructed from the resulting
displacement snapshots, and both a standard projected ROM and an ECSW HyperROM
are used for efficient online prediction.

\subsection{Numerical example: three-dimensional beam vibration}

\subsubsection{Problem setup and full-order model}

As a final benchmark, we consider a three-dimensional linear elastodynamic
problem~\cite{bhattacharyya_scikitrom_dynamic_problem1_github} on the
beam-like domain
\begin{equation}
  \Omega = [0,10]\times[0,1]\times[0,1].
\end{equation}
The domain is discretized using trilinear hexahedral elements on a structured
mesh. With the default mesh factor used in the implementation, the mesh
contains $40\times4\times4$ cells, yielding $640$ hexahedral elements. The left
face is clamped, while a time-dependent traction is applied on the right face.
The top, bottom, front, and back faces remain subject to natural boundary
conditions.

Let $\vec{u}(\vec{x},t;\bmu)$ denote the displacement field and let
$\mathbf{u}(t;\bmu)\in\mathbb{R}^{N}$ denote its finite-element coefficient
vector. The semi-discrete elastodynamic system is
\begin{equation}
  \mat{M}\ddot{\mathbf{u}}(t;\bmu)
  + \mat{C}(\bmu)\dot{\mathbf{u}}(t;\bmu)
  + \mat{K}(\bmu)\mathbf{u}(t;\bmu)
  = \mathbf{f}(t),
\end{equation}
where $\mat{M}$ is the mass matrix, $\mat{C}(\bmu)$ is the damping
matrix, $\mat{K}(\bmu)$ is the stiffness matrix, and $\mathbf{f}(t)$ is the
time-dependent load vector. The consistent mass matrix is assembled with density
\begin{equation}
  \rho = 7850,
\end{equation}
and Rayleigh damping is used:
\begin{equation}
  \mat{C}(\bmu) = c_v\,\mat{M} + c_m\,\mat{K}(\bmu),
  \qquad c_v = 10^{-3},
  \qquad c_m = 10^{-1}.
\end{equation}
Here, $c_v$ and $c_m$ are the coefficients of viscous and material damping, respectively. Time integration is performed using a Newmark scheme.

The beam is partitioned into two material regions along the $x$-direction, with
the interface at $x=5$. The constitutive response is linear elastic and
parameterized by Young's modulus $E$ and Poisson's ratio $\nu$:
\begin{equation}
  \bmu = [E,\nu]^{\top},
\end{equation}
with
\begin{equation}
  E \in [3\times10^{6},\;10^{7}], \qquad \nu \in [0.2,\;0.4].
\end{equation}
The Lam\'e parameters are computed from $(E,\nu)$ in the standard manner. To
introduce material contrast, both Lam\'e parameters are scaled by a factor of
$50$ in the left half of the beam, while the right half retains the base
material properties. The stiffness matrix is assembled through an affine
decomposition over the two material regions:
\begin{equation}
  \mat{K}(\bmu)
  = \sum_{m=1}^{2}
    \left[
      \lambda_m(E,\nu)\,\mat{K}_{m}^{(\lambda)}
      + \mu_m(E,\nu)\,\mat{K}_{m}^{(\mu)}
    \right],
\end{equation}
where $m$ indexes the material region, and $\mat{K}_{m}^{(\lambda)}$ and
$\mat{K}_{m}^{(\mu)}$ are the parameter-independent stiffness contributions
associated with the volumetric and shear parts of the elastic operator.

The external loading is a time-harmonic traction applied on the right face in
the $y$-direction:
\begin{equation}
  \vec{t}(t) =
  \begin{bmatrix}
    0 \\ -10^5\sin(3\pi t) \\ 0
  \end{bmatrix}.
\end{equation}
The simulation is carried out over the time interval
\begin{equation}
  t \in [0,\;120],
\end{equation}
using $2000$ uniformly spaced time points. Training and testing parameters are
generated using Sobol sampling. With the default choice $N_{\mathrm{snap}}=32$,
the implementation generates $32$ training samples and $32$ testing samples in
the two-dimensional parameter space $(E,\nu)$. The full-order solver returns
the displacement history over the full time window for each sampled parameter,
and these transient responses are used to construct the reduced basis.

Since the problem is linear and affine, standard Galerkin projection is
sufficient and no hyper-reduction is introduced. In the implementation, an
$r=11$ dimensional reduced model evolves a mean-subtracted displacement field.
With $\mat{U}\in\mathbb{R}^{N\times r}$ denoting the displacement basis, the
projected operators are
\begin{equation}
  \widehat{\mat{M}} = \mat{U}^{\top}\mat{M}\mat{U},
  \qquad
  \widehat{\mat{K}}(\bmu) = \mat{U}^{\top}\mat{K}(\bmu)\mat{U},
  \qquad
  \widehat{\mat{C}}(\bmu) = c_v\widehat{\mat{M}}
                              + c_m\widehat{\mat{K}}(\bmu).
\end{equation}
The reduced forcing includes both the projected external load and the mean-shift
contribution associated with the reference state used in the ROM construction.

\subsubsection{ROM results}

\begin{figure}[t]
  \centering
  \begin{subfigure}[b]{0.7\linewidth}
    \centering
    \includegraphics[width=\linewidth]{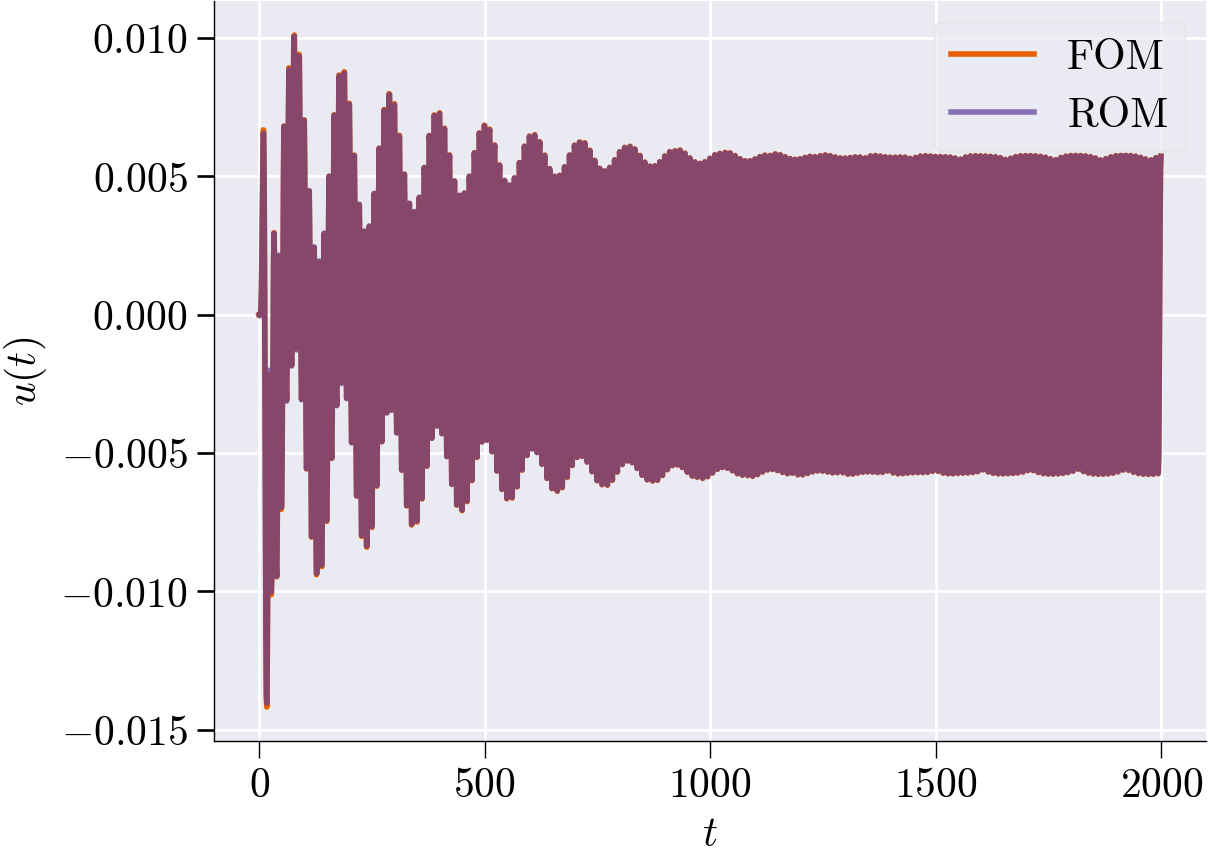}
    \caption{}
    \label{fig:ex_bv_a}
  \end{subfigure}
  \begin{subfigure}[b]{0.7\linewidth}
    \centering
    \includegraphics[width=\linewidth]{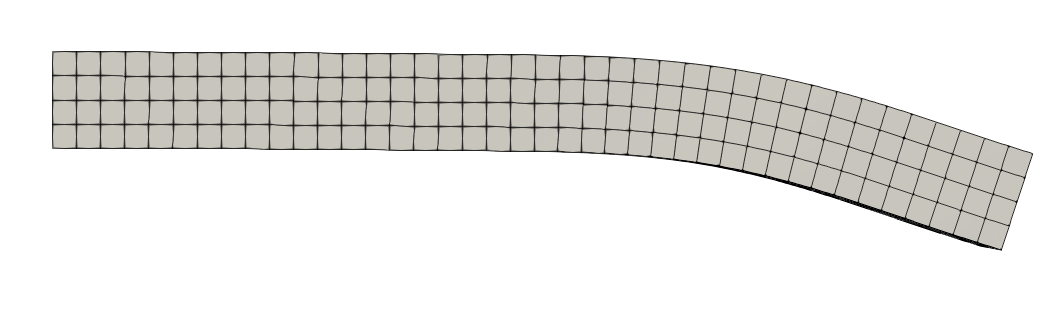}
    \caption{}
    \label{fig:ex_bv_b}
  \end{subfigure}
\caption{ROM accuracy for the three-dimensional beam vibration problem in Example~4. 
(a) Displacement history at a representative observation point; the FOM and ROM time series nearly overlap. 
(b) Representative beam deformation obtained from the FOM and ROM, where the wireframe denotes the FOM deformation and the solid surface denotes the ROM deformation.}
  \label{fig:ex4_beam_vib}
\end{figure}

\Cref{fig:ex_bv_a} compares the full-order and reduced-order displacement
histories at a representative observation point, chosen as node index~100 on
the loaded face. The 11-dimensional ROM reproduces the dominant oscillatory
response and phase evolution over the full simulation window. \Cref{fig:ex_bv_b}
shows representative deformed configurations at the fourth time step.

A quantitative summary of the ROM performance is reported in
\cref{tab:beam_vibration_results}.

\begin{table}[htbp]
  \centering
  \caption{Quantitative summary of ROM performance for the three-dimensional
    beam vibration problem in Ex.~4.}
  \label{tab:beam_vibration_results}
  \begin{tabular}{lc}
    \hline
    \textbf{Metric} & \textbf{Value} \\
    \hline
    Average speed-up                              & $35$                  \\
    Relative $L_2$ error over all parameters      & $6.1930\times10^{-3}$ \\
    Relative $L_\infty$ error over all parameters & $6.3260\times10^{-3}$ \\
    RMSE                                          & $3.2913\times10^{-4}$ \\
    $R^2$ score                                   & $1.0000$              \\
    \hline
  \end{tabular}
\end{table}

This example demonstrates the projection-based workflow of \skrom{} for
transient structural dynamics. The full-order model is solved over a
two-parameter material space, the affine stiffness structure is retained in
the reduced model, and the resulting ROM provides an efficient approximation of
the beam response under time-dependent loading.

\section{Conclusion and Future Work}
\label{sec:conclusions}

This work presented \texttt{scikit-rom}, an open-source Python library designed for projection-based reduced-order modeling (ROM) applied to finite element problems. The library delivers an intrusive ROM workflow encompassing all key stages: problem definition, high-fidelity simulations, snapshot data acquisition, basis generation, reduced operator assembly, online model application, error analysis, and hyper-reduction. All functionalities are provided in a Python environment, seamlessly built on top of \texttt{scikit-fem}.

\texttt{scikit-rom} is developed to serve not just as a computational tool, but as an accessible platform for teaching, learning, and rapid prototyping of ROM methods. Traditional projection-based ROM pipelines consist of several interlinked stages that can be challenging to dissect in an educational setting—particularly when buried behind complex software or external solvers. By integrating the full finite element model, basis construction, projection, hyper-reduction, and error metrics into a single, transparent codebase, \texttt{scikit-rom} empowers both students and researchers to follow the complete algorithmic workflow and gain insights into each component’s role in reduced modeling.

Architecturally, the library adopts a modular structure centered on problem templates, registry-based interfaces, serial and parallel driver classes, and generalized operators for both bilinear and linear forms. It accommodates not only standard Galerkin ROMs but also incorporates advanced hyper-reduction approaches—such as DEIM, S-OPT, ECSW, and ECM-style sample selection—within one unified offline-online framework. This design allows instructors to spotlight specific stages of the process, from snapshot generation and POD basis extraction to Galerkin projection, empirical interpolation, sampling strategies, reduced quadrature rules, and comparative error analysis.

To illustrate its versatility, four numerical examples were carried out across various engineering scenarios: a 1D nonlinear heat conduction case for hyper-reduction strategy comparison; a 2D transient linear heat problem with affine parameterization; a 3D finite-strain hyperelastic block utilizing ECSW; and a 3D transient beam vibration simulation. These case studies collectively address linear and nonlinear, static and transient, as well as heat transfer and solid mechanics problems. They are intended not only as benchmarks, but as interactive, modifiable learning modules, enabling users to test parameters, monitor intermediary results, compare high-fidelity and reduced solutions, and evaluate the balance of accuracy and computational savings.

The central achievement of \texttt{scikit-rom} lies in its consolidation of well-established ROM and hyper-reduction algorithms into a codebase that is concise, accessible, and conducive to both research and education. The software addresses the widespread need in computational science and engineering for readily inspectable tools—suitable for interactive notebooks, assignments, class demonstrations, and project work—that fully expose the journey from finite element models to reduced-order surrogates. In doing so, \texttt{scikit-rom} bridges the gap between ROM theory and its computational practice.

Future development will progress on multiple fronts. Planned methodological improvements include robust Petrov–Galerkin ROM capabilities, non-intrusive ROM routines, and the integration of machine learning-based hyper-reduction methods alongside existing techniques like DEIM, S-OPT, ECSW, and ECM. On the educational side, the example suite will be expanded into structured tutorials featuring guided tasks, expected solutions, and assessment elements; additional cases will broaden coverage across fluid dynamics, solid and heat transfer, and related physics. Further work will target large-scale and parallel computations, a posteriori error estimators, adaptive sampling, and performance benchmarking relative to established ROM toolkits.

To foster community engagement, reproducibility, and ease of use in classroom settings, \texttt{scikit-rom} is available via GitHub (\url{https://github.com/suparnob100/scikit-rom}) and through an accompanying tutorial website (\url{https://scikitrom.github.io}), which together provide the software, documented examples, reference material, and onboarding resources for users wishing to explore or extend projection-based ROM workflows.

\bibliographystyle{unsrtnat}
\bibliography{ref2}

\end{document}